\documentclass[journal ]{new-aiaa}
\usepackage[utf8]{inputenc}
\usepackage{textcomp}
\usepackage{graphicx}
\usepackage{amsmath}
\usepackage[version=4]{mhchem}
\usepackage{siunitx}
\usepackage{longtable,tabularx}
\usepackage{comment}
\usepackage{breqn}
\usepackage{graphicx}
\usepackage{subcaption}
\usepackage{booktabs}
\usepackage{float}
\usepackage{amsbsy}
\usepackage{bm}
\usepackage{mathtools}
\mathtoolsset{showonlyrefs}

\newcolumntype{M}[1]{>{\centering\arraybackslash}m{#1}}

\newtheorem{assumption}{Assumption}
\newtheorem{problem}{Problem}

\title{Time-Constrained Model Predictive Control for Autonomous Satellite Rendezvous, Proximity Operations, and Docking \footnote{The views expressed are those of the author and do not necessarily reflect the official policy or position of the Department of the Air Force, the Department of Defense, or the U.S. government. Approved for public release; distribution is unlimited. Public Affairs approval \#AFRL20250218.}}

\author{Gabriel Behrendt, \footnote{Ph.D. Candidate, School of Electrical and Computer Engineering} Matthew Hale \footnote{Associate Professor, School of Electrical and Computer Engineering} }
\affil{Georgia Institute of Technology, Atlanta, GA, 30332}
\author{Alexander Soderlund, \footnote{Research Aerospace Engineer, Space Control Branch, AIAA Member} Sean Phillips, \footnote{Technology Advisor, Space Control Branch, AIAA Member} Evan Kain \footnote{Computer Engineer, Spacecraft Component Technology}}
\affil{Air Force Research Laboratory, Kirtland AFB, NM, 87117}

\begin{document}
\newcommand{\alex}[1]{\textcolor{green}{#1}}
\newcommand{\gb}[1]{\textcolor{blue}{#1}}
\newcommand{\todo}[1]{\textcolor{red}{\underline{\bf TODO}: #1}}

\newcommand{\reals}{\mathbb{R}}
\newcommand{\sphere}{\mathbb{S}}
\newcommand{\nnreals}{\mathbb{R}_{\geq 0}}
\newcommand{\R}{\mathbb{R}}
\newcommand{\ones}{\textbf{1}}

\newcommand{\X}{\mathcal{X}}
\newcommand{\U}{\mathcal{U}}
\newcommand{\D}{\mathcal{D}}

\newcommand{\ints}{\mathbb{Z}}
\newcommand{\nnints}{\mathbb{Z}_{\geq 0}}

\newcommand{\statespace}{\mathcal{X}}
\newcommand{\inputspace}{\mathcal{U}}

\newcommand{\statemat}{A}
\newcommand{\inputmat}{B}

\newcommand{\dt}{\Delta t}
\newcommand{\discstatemat}{A_{dt}}
\newcommand{\discinputmat}{B_{dt}}

\newcommand{\state}{\mathbf{x}}
\newcommand{\sysinput}{\mathbf{u}}

\newcommand{\traj}{\xi}
\newcommand{\initstate}{\theta}

\newcommand{\disctraj}{\xi_{dt}}
\newcommand{\stattransmat}{\phi}

\newcommand{\nmt}{\mathcal{M}}

\newcommand{\eulang}{\Gamma}
\newcommand{\roll}{\varphi}
\newcommand{\pitch}{\theta}
\newcommand{\yaw}{\psi}
\newcommand{\angularvel}[3]{\omega_{\mathcal{#1}\mathcal{#2}}^{\mathcal{#3}}}
\newcommand{\dotangularvel}[3]{\dot{\omega}_{\mathcal{#1}\mathcal{#2}}^{\mathcal{#3}}}
\newcommand{\angularvelerr}[3]{\delta\omega_{\mathcal{#1}\mathcal{#2}}^{\mathcal{#3}}}
\newcommand{\dotangularvelerr}[3]{\delta\dot{\omega}_{\mathcal{#1}\mathcal{#2}}^{\mathcal{#3}}}
\newcommand{\angularveldes}[3]{\tilde{\omega}_{\mathcal{#1}\mathcal{#2}}^{\mathcal{#3}}}
\newcommand{\dotangularveldes}[3]{\dot{\tilde{\omega}}_{\mathcal{#1}\mathcal{#2}}^{\mathcal{#3}}}
\newcommand{\angularmom}[1]{h_{cm}^{\mathcal{#1}}}
\newcommand{\dotangularmom}[1]{\dot{h}_{cm}^{\mathcal{#1}}}

\newcommand{\angularaccel}[3]{\alpha_{\mathcal{#1}\mathcal{#2}}^{\mathcal{#3}}}

\newcommand{\torque}{\tau}

\newcommand{\inertia}[1]{J^{\mathcal{#1}}}
\newcommand{\dotinertia}[1]{\dot{J}^{\mathcal{#1}}}
\newcommand{\inertiainv}[1]{K^{\mathcal{#1}}}

\newcommand{\specorg}[1]{\relax\ifmmode 
{\mathbf{SO}}(#1)
\else $\mathbf{SO}(#1)$
\fi}
\newcommand{\speceul}[1]{\relax\ifmmode 
{\mathbf{SE}}(#1)
\else $\mathbf{SE}(#1)$
\fi}

\newcommand{\so}[1]{\relax\ifmmode 
{\mathfrak{so}}(#1)
\else $\mathfrak{so}(#1)$
\fi}

\newcommand{\se}[1]{\relax\ifmmode 
{\mathfrak{se}}(#1)
\else $\mathfrak{se}(#1)$
\fi}

\newcommand{\skewmat}[1]{\relax\ifmmode 
{\mathcal{S}}(#1)
\else $\mathcal{S}(#1)$
\fi}

\newcommand{\skewop}[1]{\relax\ifmmode 
[#1]^{\times}
\else $[#1]^{\times}$
\fi}

\newcommand{\coordframe}[1]{\mathcal{#1}}
\newcommand{\identity}[1]{I_{#1}}
\newcommand{\rotmat}[2]{\relax\ifmmode 
{R_{\mathcal{#1}}^{\mathcal{#2}}}
\else $R_{\mathcal{#1}}^{\mathcal{#2}}$
\fi}

\newcommand{\multiGauss}[2]{\mathcal{N}(#1, #2)}

\newcommand{\dotrotmat}[2]{\dot{R}_{\mathcal{#1}}^{\mathcal{#2}}}
\newcommand{\quat}[2]{q_{\mathcal{#1}}^{\mathcal{#2}}}
\newcommand{\dotquat}[2]{\dot{q}_{\mathcal{#1}}^{\mathcal{#2}}}
\newcommand{\quatscalar}[2]{\eta_{\mathcal{#1}}^{\mathcal{#2}}}
\newcommand{\quatvector}[2]{\rho_{\mathcal{#1}}^{\mathcal{#2}}}
\newcommand{\dotquatscalar}[2]{\dot{\eta}_{\mathcal{#1}}^{\mathcal{#2}}}
\newcommand{\dotquatvector}[2]{\dot{\rho}_{\mathcal{#1}}^{\mathcal{#2}}}
\newcommand{\quaterr}[2]{\delta q_{\mathcal{#1}}^{\mathcal{#2}}}
\newcommand{\dotquaterr}[2]{\delta\dot{q}_{\mathcal{#1}}^{\mathcal{#2}}}
\newcommand{\quatscalarerr}[2]{\delta\eta_{\mathcal{#1}}^{\mathcal{#2}}}
\newcommand{\quatvectorerr}[2]{\delta\rho_{\mathcal{#1}}^{\mathcal{#2}}}
\newcommand{\dotquatscalarerr}[2]{\delta\dot{\eta}_{\mathcal{#1}}^{\mathcal{#2}}}
\newcommand{\dotquatvectorerr}[2]{\delta\dot{\rho}_{\mathcal{#1}}^{\mathcal{#2}}}
\newcommand{\quatdes}[2]{\tilde{q}_{\mathcal{#1}}^{\mathcal{#2}}}
\newcommand{\dotquatdes}[2]{\dot{\tilde{q}}_{\mathcal{#1}}^{\mathcal{#2}}}
\newcommand{\quatscalardes}[2]{\tilde{\eta}_{\mathcal{#1}}^{\mathcal{#2}}}
\newcommand{\quatvectordes}[2]{\tilde{\rho}_{\mathcal{#1}}^{\mathcal{#2}}}
\newcommand{\dotquatscalardes}[2]{\dot{\tilde{\eta}}_{\mathcal{#1}}^{\mathcal{#2}}}
\newcommand{\dotquatvectordes}[2]{\dot{\tilde{\rho}}_{\mathcal{#1}}^{\mathcal{#2}}}

\newcommand{\norm}[1]{||#1||}
\newcommand{\jmax}{j_\max}

\maketitle
\begin{abstract}
This paper presents a time-constrained model predictive control strategy for the six degree-of-freedom autonomous rendezvous, proximity, operations and docking problem between a controllable ``deputy'' satellite and an uncontrolled ``chief'' satellite. The objective is to achieve a docking configuration defined by both the translational and attitudinal states of the deputy relative to the chief, whose dynamics are respectively governed by both the Clohessy-Wiltshire equations and Euler's second law of motion. The proposed control strategy explicitly addresses computational time constraints that are common to state-of-the-art space vehicles. Thus, a time-constrained model predictive control strategy is implemented on a space-grade processor. Although suboptimal with regards to energy consumption when compared to conventional optimal RPO trajectories, it is empirically demonstrated via numerical simulations that the deputy spacecraft still achieves a successful docking configuration while subject to computational time constraints.
\end{abstract}




\section{Introduction}
Autonomy within the space domain is becoming increasingly necessary due to the growing number of satellites in orbit and a lack of ground-based human operators to control them. Thus, the autonomous rendezvous, proximity operations, and docking (ARPOD) problem has gained interest due to missions such as the NASA Demonstration of Autonomous Rendezvous Technology (DART)~\cite{sarli2017nasa}, the Engineering Test Satellite VII (ETS-VII)~\cite{kawano1999result}, and more recently the SpaceX Cargo Dragon using autonomous capabilities to dock with the International Space Station~\cite{moran2021}.
Furthermore, the ARPOD problem is relevant in applications such as autonomous satellite inspection~\cite{bernhard2020coordinated}, on-orbit servicing~\cite{flores2014review}, and refueling~\cite{salazar2007high} to extend satellite life and reduce costs.

In general, the ARPOD problem requires a deputy spacecraft to autonomously maneuver to some desired state relative to a chief spacecraft. On a high-level, the deputy must complete a recursive three-step process: (i) sense its surrounding environment; (ii) update its internal state relative to that environment; and (iii) generate and perform some control action based on its state to drive the system towards the docking configuration. It is this third step that we investigate in this work, as there are significant challenges that must be addressed in the ARPOD problem in order to design a 
reliable and practically-implementable control strategy. Namely, the dynamics of the satellites are nonlinear when both the translational and attitude dynamics are considered together. And while there are a myriad of nonlinear control protocols available, it must be stressed that this introduces a significant challenge, as these control inputs must be computed \textit{locally}, i.e. on-board a space-grade processor housed within the spacecraft bus. 
Since space-grade processors must be radiation-hardened (which requires lengthy construction cycles) to handle the electronics-hostile space environment \cite{bourdarie2008near} and the design of space missions are often multi-year endeavors, there is a sizable processing capabilities gap between state-of-the-art, commercial off-the-shelf CPUs and their space-grade counterparts \cite{lovelly2017comparative, lovelly2017comparative_dissertation}. This gap is expected to become more drastically illuminated in the coming decades due to two compounding factors: (i) the number of deployed space assets (and by extension spacecraft-to-spacecraft interactions) is accelerating \cite{gaston2023environmental} and (ii) space-based are becoming multifunctional and increasingly complex, particularly in the congested low-Earth orbit domain. Thus, we require a control strategy that can handle nonlinearities in the underlying dynamics whilst simultaneously executing under computational constraints. 

Autonomous control problems are often solved via state feedback methods. Model predictive control (MPC) has been widely utilized \cite{falcone2007} and is a natural fit for the ARPOD problem due to its robustness and ability to handle nonlinearities. In conventional MPC, an optimization problem is solved at each time step to completion, i.e., 
a new control input is computed by solving an optimization problem until a stopping condition is reached. However, given the limited computational capabilities that are available in space, we cannot reliably use conventional MPC because new inputs to the system may be needed before the computations to find those inputs can be completed. 
Therefore, in this work, we use a method that we refer to as \emph{time-constrained MPC}. We consider a setting in which the underlying optimization algorithm is only allowed enough time to complete a limited number of iterations when computing each input. This constraint only allows the algorithm to make some progress toward the optimum, and, when the time constraint is reached, a sub-optimal input is applied to the system by using the optimization algorithm's most recent iterate. 
It is known in the MPC literature that the optimization problem does not need to be solved to completion in order to ensure stability of the solution~\cite{mayne2014,graichen2012, graichen2010, pavlov2019}.
These works suggest that time-constrained MPC is a viable solution to account for nonlinear dynamics and limited onboard computational speed seen in the ARPOD problem, and this paper formalizes and confirms this point. 

The ARPOD problem has been studied in a variety of settings. Some works have considered only controlling translational~\cite{jewison2017guidance,chu2016optimised} or attitudinal dynamics~\cite{leomanni2014,trivailo2009}. Typically, ARPOD studies consider translational control inputs that are uncoupled from the attitudinal dynamics, such as in \cite{hogan2014attitude}. While recent studies \cite{soderlund2021autonomous, soderlund2022switching} have considered coupled translational-attitudinal systems, only two-dimensional motion between the deputy and chief was considered. Other works have developed control strategies for the more complex three-dimensional 6DOF ARPOD problem with coupled translational and attitude dynamics. These works include approaches from nonlinear control such as back-stepping \cite{sun2015, wang2019}, sliding mode~\cite{yang2019, zhou2020robust},  learning-based control~\cite{hu2021}, and artificial potential functions~\cite{dong2018}. 
This paper differs from those earlier works in that we do not use a static control law with user-defined gains. Instead, we use time-constrained MPC to solve for our next control input. 
Other works have considered optimization based control strategies.
\cite{ventura2017fast} proposed a 6-DOF trajectory generation algorithm by sequential quadratic programming where the chaser's states are represented as polynomials in the optimal control problem to reduce computational speed. 
\cite{zhou2019receding} proposed a receding horizon implementation of sequential convex programming for the
spacecraft 6-DOF ARPOD problem by linearizing the dynamics.
This paper differs by considering a nonlinear dynamic model without any polynomial or linear approximation and we explicitly address computational time constraints.

MPC for autonomous rendezvous has been used considering only translational states, i.e., relative position and velocity~\cite{hartley2015tutorial,fear2024autonomous,di2012model,jewison2015model}.
MPC has been implemented for the 6-DOF for autonomous planetary landing using piecewise affine model approximation~\cite{lee2017constrained}.
These problems can be efficiently solved by convex optimization solvers and implemented onboard due to the convexity of the problem when considering linear or affine dynamic constraints.
However, in this work we do not consider any approximation to the nonlinear dynamic model which causes the MPC problem to be more difficult to solve and take more time.
Speed ups to conventional MPC have been proposed by transforming the MPC solution  into a pre-computed lookup table~\cite{malyuta2019approximate} and custom predictive controller hardware~\cite{hartley2013graphical}.
The most closely related work to the current article is~\cite{pong2011autonomous} where the authors simulate MPC for the 6 DOF ARPOD problem offline and implement the solution in an open loop fashion. The authors note that conventional MPC was not implementable on their satellites due to limited onboard processing capabilities. 
The current article addresses this point by explicitly addressing computational time constraints in the time-constrained MPC formulation and simulating it on a satellite processing unit.
Furthermore, in our previous work~\cite{behrendt2023autonomous} we implemented time-constrained MPC on a~\emph{terrestrial-grade computer} for one initial condition. The current article extends our previous work by implementing time-constrained MPC on a SpaceCloud iX10-101 processing unit for 200 initial conditions.
To summarize, this paper makes the following contributions: 
\begin{itemize}
    \item The six degree of freedom ARPOD problem was modeled whilst considering both translational and attitudinal dynamics (Section~\ref{sec:dynamics}).
    \item Time-constrained MPC for the 6-DOF ARPOD problem was implemented on a satellite processing unit (Section~\ref{sec:hardware})
    \item A successful docking configuration was empirically validated via numerical siumlations (Section~\ref{sec:results})
\end{itemize}
The rest of the paper is organized as follows:
Section~\ref{sec:prelim} presents the mathematical preliminaries and background information related to the ARPOD problem.
Section~\ref{sec:dynamics} gives the relative motion dynamics for the 6 DOF ARPOD problem.
Then, Section~\ref{sec:control} provides the time-constrained MPC strategy which we implement on a space-grade processing unit in Section~\ref{sec:simulation}. Concluding remarks are given in Section~\ref{sec:conclusion}.

\section{Preliminaries} \label{sec:prelim}
The Euclidean $n$-dimensional space is denoted by $\reals^{n}$ and the set of natural numbers is denoted by~$\mathbb{N} = \{0,1,2,...\}$. The vector $v \in \reals^{n}$ is defined as $v := (v_1, \dots, v_n)^{\top}$, where the superscript $\top$ denotes the transpose operation. Unless otherwise specified, all vectors used in this paper are \textit{physical} vectors, meaning that they exist irrespective of any coordinate frame used. For instance, the vector $v^{\mathcal{A}}$ denotes that the vector is given by the coordinates defined by the frame $\mathcal{A}$, while $v^{\mathcal{B}}$ denotes that the vector is given by the coordinates defined by the frame $\mathcal{B}$. For~$v\in\R^3$,~$\skewop{v}$ denotes the skew-symmetric operator, i.e.,
\begin{equation}
    \skewop{v} = \begin{bmatrix}
                0 & -v_3 & v_2 \\
                v_3 & 0 & -v_1 \\
                -v_2 & v_1 & 0
                \end{bmatrix}.
\end{equation}
 The $n \times n$ identity matrix is denoted by $\identity{n}$, while $\textbf{0}_{n}$ denotes the zero column vector with dimension~$n\times 1$. 
  Similarly, the~$n \times 1$-dimensional column 
  vector of ones is denoted by~$\ones_n$. 

The Lie Group \specorg{3} is the set of all real invertible $3 \times 3$ matrices that are orthogonal with determinant $1$, i.e.,
$$\specorg{3} := \{\rotmat{}{} \in \reals^{3 \times 3} \ | \ \rotmat{}{}^{\top}\rotmat{}{} = \identity{3}, \det[\rotmat{}{}] = 1\}. $$
In this work, elements of $\specorg{3}$ will be referred to as \textit{rotation matrices}. The coordinate transformation from frame $\coordframe{B}$ to frame $\coordframe{A}$ is given by
$v^{\mathcal{A}} = \rotmat{\mathcal{B}}{\mathcal{A}} v^{\mathcal{B}}$.
Any rotation matrix $\rotmat{}{}$ can be parameterized through the unit quaternion vector
$\quat{}{} := (\quatscalar{}{},\quatvector{}{\top} )^{\top} ,$
where $\quatvector{}{} = (\rho_1, \ \rho_2, \ \rho_3)^{\top} \in \reals^3$ is the vector part and $\quatscalar{}{} \in \R$ is the scalar part. The unit quaternion is contained in the unit hypersphere, $\sphere^3$, defined with four coordinates $s := (s_1 \ s_2 \ s_3 \ s_4)^{\top} \in \reals^4$ as $\sphere^3 := \{s \in \reals^4 : s^{\top}s = 1\}.$
The inverse quaternion is 
$q^{-1} = ( \quatscalar{}{} \ {-\rho}^{\top})^{\top}$
and the identity rotation is 
$q^{I} = ( 1 \ \mathbf{0}_3^{\top} )^{\top}.$

We use $\quat{B}{A}$ to denote the quaternion corresponding to the rotation matrix $\rotmat{B}{A}$. The mapping from a quaternion $\quat{}{} = (\eta \ \rho^\top)^{\top}$ to its rotation matrix $\rotmat{}{}$ is
\begin{equation}
\label{eq:quat2rot}
\identity{3} - 2\quatscalar{}{}\skewop{\quatvector{}{}} + 2\skewop{\quatvector{}{}}\skewop{\quatvector{}{}}.
\end{equation}
We note that a quaternion $\quat{}{} = (\eta \ \rho^\top)^{\top}$ and its negative $-\quat{}{}$ represent the same physical rotation\footnote{This work has opted to adopt the convention of \citep{kuipers1999quaternions} where the passive rotation operator on a vector $v$ is represented as $q^{-1} \otimes v \otimes q$ where $v$ is a ``pure" quaternion with $0$ scalar part. This choice affects the signs of Eq.\eqref{eq:quat2rot} and the angular velocity kinematics.}.

This work denotes $\angularvel{\coordframe{E}}{\coordframe{A}}{\coordframe{B}} \in \reals^{3}$ as the angular velocity of frame $\coordframe{A}$ \textit{relative} to frame $\coordframe{E}$ but with vector components represented in frame $\coordframe{B}$. The time rate change of any rotation matrix $\rotmat{\mathcal{B}}{\mathcal{A}}$ is given as
\begin{align}
\label{eq:rotation_velocity_a}
\dotrotmat{\mathcal{B}}{\mathcal{A}} &= \rotmat{\mathcal{B}}{\mathcal{A}}\skewop{\angularvel{\coordframe{A}}{\coordframe{B}}{\coordframe{B}}} \\
\label{eq:rotation_velocity_b}
\dotrotmat{\mathcal{B}}{\mathcal{A}} &= \skewop{\angularvel{\coordframe{A}}{\coordframe{B}}{\coordframe{A}}}\rotmat{\mathcal{B}}{\mathcal{A}}.
\end{align}
Analogously, the kinematics of a quaternion $\quat{A}{B} = (\quatscalar{}{} \ \rho^{\top})^{\top}$ are related to the angular velocity as
\begin{align}
\label{eq:quaternion_velocity_c}
\begin{pmatrix} \dotquatscalar{}{} \\ \dotquatvector{}{} \end{pmatrix} &= 
-\frac{1}{2}\begin{pmatrix}
-\rho^{\top} \\ \eta\identity{3} + \skewop{\rho}
\end{pmatrix} \angularvel{\coordframe{A}}{\coordframe{B}}{\coordframe{A}}.
\end{align}
\section{Relative Motion Dynamics for ARPOD} \label{sec:dynamics}
\begin{figure}[ht]
\centering
\includegraphics[width=0.35\textwidth]{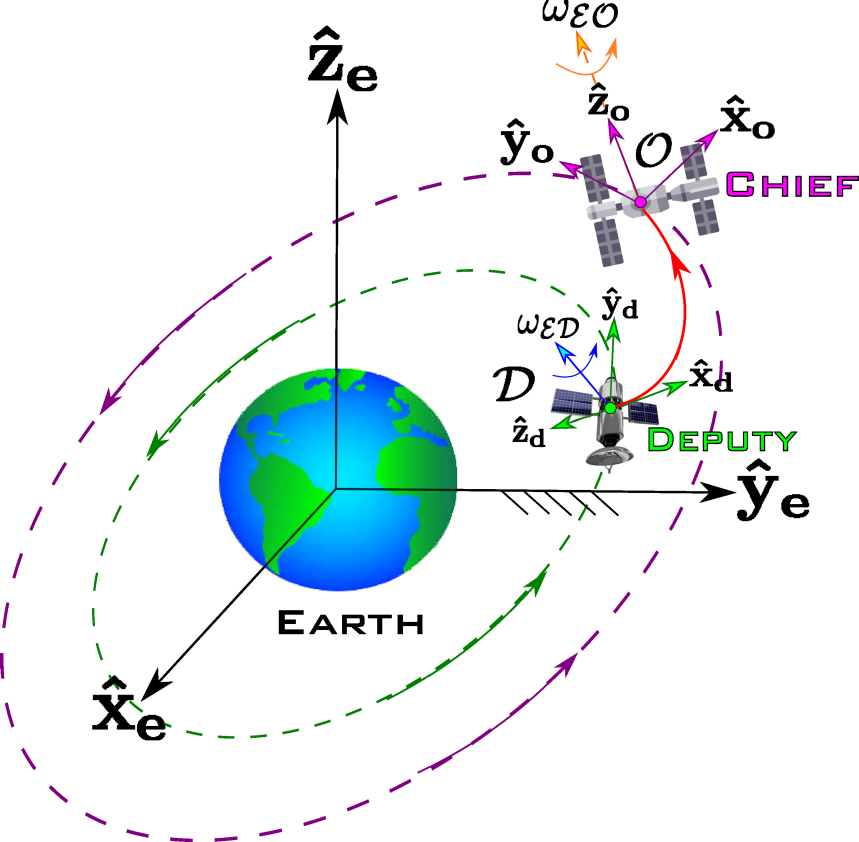}
\caption{The chief (with rotating orbit-fixed frame $\coordframe{O}$) and a deputy (with body-fixed frame $\coordframe{D}$) are orbiting about the Earth with inertial frame $\coordframe{E}$. The dashed lines are the closed orbital trajectories of both spacecraft. The red solid line depicts the rendezvous trajectory from the deputy to the chief as seen in the inertial frame.}
\label{fig:ExclusionZones}
\end{figure}

The objective of this work is to design an autonomous controller such that a deputy spacecraft's translational and attitudinal states converge to a successful docking configuration with the chief, as shown in Figure~\ref{fig:ExclusionZones}. 
Three frames of reference are used: 
\begin{enumerate}
\item {The Earth-Centered Inertial (ECI) frame originating from the Earth's center is denoted $\coordframe{E} := \{\mathbf{\hat{x}_e}, \mathbf{\hat{y}_e}, \mathbf{\hat{z}_e}\}$.}
\item {The Clohessy-Wiltshire (CW) frame is a non-inertial frame attached to the chief satellite, denoted $\mathcal{O} := \{\mathbf{\hat{x}_o}, \mathbf{\hat{y}_o}, \mathbf{\hat{z}_o}\}$. The origin lies at the center of mass of the chief, unless otherwise specified. The $\mathbf{\hat{x}_o}$-axis aligns with the vector pointing from the center of the Earth towards the center of mass of the chief satellite. The $\mathbf{\hat{z}_o}$-axis is in the direction of the orbital angular momentum vector of the chief spacecraft. The $\mathbf{\hat{y}_o}$-axis completes the right-handed orthogonal frame.}
\item{The deputy-fixed frame is a non-inertial body-fixed frame attached to the deputy satellite and denoted $\mathcal{D} := \{\mathbf{\hat{x}_d}, \mathbf{\hat{y}_d}, \mathbf{\hat{z}_d}\}$. The origin is assumed to lie at the center of mass of the deputy, unless otherwise specified. We use the convention that the $\mathbf{\hat{x}_d}$-axis is along the starboard direction, $\mathbf{\hat{y}_d}$ is in the fore direction, and $\mathbf{\hat{z}_d}$ completes the right-handed frame in the topside direction.}
\end{enumerate}
Let $(m_c, \inertia{C})$ and $(m_d, \inertia{D})$ be the mass-inertia pairs of the chief and deputy respectively. We make the following assumptions regarding their relative motion dynamics:

\begin{assumption} \label{as:A1}
    Both the chief and the deputy spacecraft are assumed to be rigid bodies of constant mass
\end{assumption}
\begin{assumption} \label{as:A2}
    The chief-fixed frame is assumed to always align with the orbit-fixed frame $\coordframe{O}$.
\end{assumption}
\begin{assumption} \label{as:A3}
    The chief spacecraft is in an uncontrolled, circular orbit.
\end{assumption}
\begin{assumption} \label{as:A4}
    The distance between both satellites is much less than the distance between the Earth and the chief.
\end{assumption}
\begin{assumption} \label{as:A5}
    The deputy spacecraft has bi-directional thrusters and external torque-generators installed along the axes aligning with its body-fixed frame.
\end{assumption}
\begin{assumption} \label{as:A6}
    The axes of the deputy-fixed frame are aligned with the principal axes of the deputy body.
\end{assumption}

\subsection{Translational Dynamics in the Chief Frame}
 The position and velocity of the deputy relative to the chief with respect to the CW frame $\coordframe{O}$ are
\begin{equation}
\delta r^{\coordframe{O}} := (\delta x \ \delta y \ \delta z)^{\top} 
\textnormal{ and }
\delta \dot{r}^{\coordframe{O}} := (\delta \dot{x} \ \delta \dot{y} \ \delta \dot{z})^{\top}.
\end{equation}
Assumptions~\ref{as:A3}-\ref{as:A5} allow the use of the Clohessy-Wiltshire translational dynamics (see \citep{curtis2013orbital}) with deputy controls, expressed as
\begin{equation}
\label{eq:CWdynamics_translational}
\begin{pmatrix}
\delta \dot{x} \\ \delta \dot{y} \\ \delta \dot{z} \\ \delta \ddot{x} \\ \delta \ddot{y} \\ \delta \ddot{z}
\end{pmatrix}
= \begin{pmatrix}
0 & 0 & 0 & 1 & 0 & 0 \\
0 & 0 & 0 & 0 & 1 & 0 \\
0 & 0 & 0 & 0 & 0 & 1 \\
3n^2& 0 & 0 & 0 & 2n & 0 \\
 0 & 0 & 0 & -2n & 0 & 0 \\
 0 & 0 & -n^2 & 0 & 0 & 0 
\end{pmatrix}
\begin{pmatrix}
\delta x \\ \delta y \\ \delta z \\ \delta \dot{x} \\ \delta \dot{y} \\ \delta \dot{z}
\end{pmatrix} + 
\begin{pmatrix}
\mathbf{0}_{3 \times 3}\\
\frac{1}{m_d}\rotmat{D}{O}
\end{pmatrix}
F^{\coordframe{D}}_d,
\end{equation}
 where the ``mean motion" constant is $n = \sqrt{\frac{\mu}{\norm{r_c}^3}}$ in $\text{rad}\cdot \text{s}^{-1}$, $\mu = 398600.4418 \ \text{km}^{3}\text{s}^{-2}$ is the Earth's standard gravitational parameter,~$r_c$ is the radius of the chief's circular orbit, and~$\rotmat{D}{O}$ is the rotation matrix applied to~$F^{\coordframe{D}}_d$ to represent the deputy's thrust vector in the Chief frame.

\subsection{Attitudinal Dynamics in the Chief Frame} \label{sec:new_rotation_dynamics}

Both frames $\coordframe{O}$ and $\coordframe{D}$ are rotating with respect to the inertial frame $\coordframe{E}$. The angular velocity equations are
$\angularvel{E}{D}{O} := (\omega_1 \ \omega_2 \ \omega_3)^{\top}$,  
$\angularvel{O}{D}{O} := (\delta\omega_1 \ \delta\omega_2 \ \delta\omega_3)^{\top}$,
and
$\angularvel{E}{O}{O} := (0 \ 0 \ n)^{\top}$. 
The deputy external torque vector in the $\coordframe{O}$ frame is $\tau^{\coordframe{O}} := (\tau_1 \ \tau_2 \ \tau_3)^{\top}$ and from Assumption~\ref{as:A6} the deputy inertia matrix in the deputy-fixed frame is $$\inertia{D} = \begin{pmatrix}
J_1 & 0 & 0 \\ 0 & J_2 & 0 \\ 0 & 0 & J_3
\end{pmatrix}.$$

The angular velocity of the $\coordframe{D}$ frame relative to the $\coordframe{O}$ frame is 
$\angularvel{O}{D}{E} = \angularvel{E}{D}{E} - \angularvel{E}{O}{E}$,
with the time derivatives given by 
\begin{equation}
\label{eq:rel_ang_vel_inertial}
\dotangularvel{O}{D}{E} = \dotangularvel{E}{D}{E} - \dotangularvel{E}{O}{E}.
\end{equation}
From \eqref{eq:rotation_velocity_a} and \eqref{eq:rotation_velocity_b} we yield the relations 
\begin{align*}
\dotangularvel{O}{D}{E} &= \rotmat{O}{E}\skewop{\angularvel{E}{O}{O}}\angularvel{O}{D}{O} + \rotmat{O}{E}\dotangularvel{O}{D}{O} \\
\dotangularvel{E}{O}{E} &= \rotmat{O}{E}\underbrace{\skewop{\angularvel{E}{O}{O}}\angularvel{E}{O}{O}}_{=0} + \rotmat{O}{E}\underbrace{\dotangularvel{E}{O}{O}}_{=0} = 0  \\
\dotangularvel{E}{D}{E} &= \rotmat{D}{E}\underbrace{\skewop{\angularvel{E}{D}{D}}\angularvel{E}{D}{D}}_{=0} + \rotmat{D}{E}\dotangularvel{E}{D}{D},
\end{align*}
which, when combined with~\eqref{eq:rel_ang_vel_inertial}, gives
the relation 
$\dotangularvel{O}{D}{O} =  \rotmat{D}{O}\dotangularvel{E}{D}{D} - \skewop{\angularvel{E}{O}{O}}\angularvel{O}{D}{O}$.
Let the deputy body have inertia matrix $\inertia{D}$ as measured in the deputy-fixed frame with external control torque vector $\tau^{\coordframe{D}}$ being applied about the center of mass. From Euler's second law of motion we have
\begin{equation}
\label{eq:deputy_slew}    
\dotangularvel{E}{D}{D} = -\inertiainv{D}\skewop{\angularvel{E}{D}{D}}\inertia{D}\angularvel{E}{D}{D} + \inertiainv{D}\tau^{\coordframe{D}},
\end{equation}
where $\inertiainv{D} := [\inertia{D}]^{-1}$. Application of rotational transformation to \eqref{eq:deputy_slew} gives
$$\dotangularvel{O}{D}{O} =  \rotmat{D}{O}\left( -\inertiainv{D}\skewop{\angularvel{E}{D}{D}} \inertia{D}\angularvel{E}{D}{D} + \inertiainv{D}\tau^{\coordframe{D}} \right) - \skewop{\angularvel{E}{O}{O}}\angularvel{O}{D}{O},$$
which expands to
\begin{dmath}
\label{eq:ugly_ass_equation}
\dotangularvel{O}{D}{O} = \skewop{\angularvel{O}{D}{O}}\angularvel{E}{O}{O} + \inertiainv{O} \tau^{\coordframe{O}} -  \inertiainv{O} \Big(\skewop{\angularvel{O}{D}{O}}
\inertia{D}\rotmat{O}{D}\angularvel{O}{D}{O} + \skewop{\angularvel{O}{D}{O}}
\inertia{D}\rotmat{O}{D}\angularvel{E}{O}{O} + \skewop{\angularvel{E}{O}{O}}
\inertia{D}\rotmat{O}{D}\angularvel{O}{D}{O} + \skewop{\angularvel{E}{O}{O}}
\inertia{D}\rotmat{O}{D}\angularvel{E}{O}{O} \Big).
\end{dmath}
where $\inertiainv{O} = \rotmat{D}{O}\inertiainv{D}\rotmat{O}{D}$.

\subsection{Docking Configuration} \label{sec:docking_config}
To achieve a successful docking configuration, the deputy must reach a set of predefined relative translational and attitudinal states. Let $\quatdes{D}{O}$ be the quaternion representing the desired rotation from the deputy $\coordframe{D}$ frame to the CW $\coordframe{O}$ frame, and let $\quat{D}{O}$ be the actual (i.e., plant-state) rotation quaternion. Analogously, let $\angularveldes{O}{D}{O}$ be the desired relative angular velocity and $\angularvel{O}{D}{O}$ be the actual relative angular velocity. Define $\quaterr{D}{O} := \{\quatdes{D}{O}\}^{-1}\otimes\quat{D}{O}$ and $\angularvelerr{O}{D}{O} := \angularvel{O}{D}{O} - \angularveldes{O}{D}{O}$ which respectively denote the error quaternion and error angular velocity. 
To achieve on-orbit attitudinal synchronization between both spacecraft, the aim is to drive $\quaterr{D}{O} \to q^{I}$ and $\angularvelerr{O}{D}{O} \to \mathbf{0}_3$. 
In the ARPOD problem, $\quatdes{D}{O} = q^{I}$ and $\angularveldes{O}{D}{O} = \mathbf{0}_3$, which results in the attitudinal error kinematics
\begin{align} 
    \dotquatscalarerr{D}{O} &= \frac{1}{2}[\quatvectorerr{D}{O}]^{\top}\angularvelerr{O}{D}{O} \label{eq:scalarQuatDynamics} \\
    \dotquatvectorerr{D}{O} &= -\frac{1}{2}[\quatscalarerr{D}{O}\identity{3} + \skewop{\quatvectorerr{D}{O}}]\angularvelerr{O}{D}{O}. \label{eq:vectorQuatDynamics}
\end{align}
Relative translational states required for docking are $\delta r^{\coordframe{O}} = \textbf{0}_{3}$ and $\delta \dot{r}^{\coordframe{O}} = \textbf{0}_{3}$. We combine the translational and attitudinal states to form the state vector
\begin{equation}
x \coloneqq \big([\delta r^{\coordframe{O}}]^{\top}, \ [\delta\dot{r}^{\coordframe{O}}]^{\top}, \ \quatscalarerr{D}{O}, \ [\quatvectorerr{D}{O}]^{\top}, \ [\angularvelerr{O}{D}{O}]^{\top}\big)^{\top} \in \R^{13}.
\end{equation}
The desired docking state is
\begin{equation}
\label{eq:docking}
x_d \coloneqq  \left( [\mathbf{0}_3]^{\top}, \ [\mathbf{0}_3]^{\top}, \ 1, \ [\mathbf{0}_3]^{\top}, \ [\mathbf{0}_3]^{\top} \right)^{\top} .
\end{equation}
This vector~$x_d$ 
indicates that the frame~$\coordframe{O}$ has the same origin, linear velocity, orientation, and angular velocity as frame~$\coordframe{D}$. 
Let the control vector of interest in frame $\coordframe{D}$ be~$u \coloneqq  \big( [F^{\coordframe{D}}_d]^{\top}, \ [\tau^{\coordframe{D}}]^{\top} \big)^{\top} \in \R^6$
and the desired docking inputs be~$u_d \coloneqq  \big( \mathbf{0}_3^{\top}, \ \mathbf{0}_3^{\top} \big)^{\top} \in \R^6$.
This desired input vector~$u_d$ indicates the deputy is not applying thrusts or torques about its axes when the docking configuration is achieved.
Lastly, let us define the combined state and input vector~$z \coloneqq \big( u^\top, \ x^\top \big)^\top \in \R^{19}$ and the desired docking configuration~$z_d \coloneqq \big( u_d^\top, \ x_d^\top \big)^\top \in \R^{19}$.

\section{Control Strategy} \label{sec:control}

\begin{figure}[H]
\centering
\includegraphics[width=0.6\textwidth]{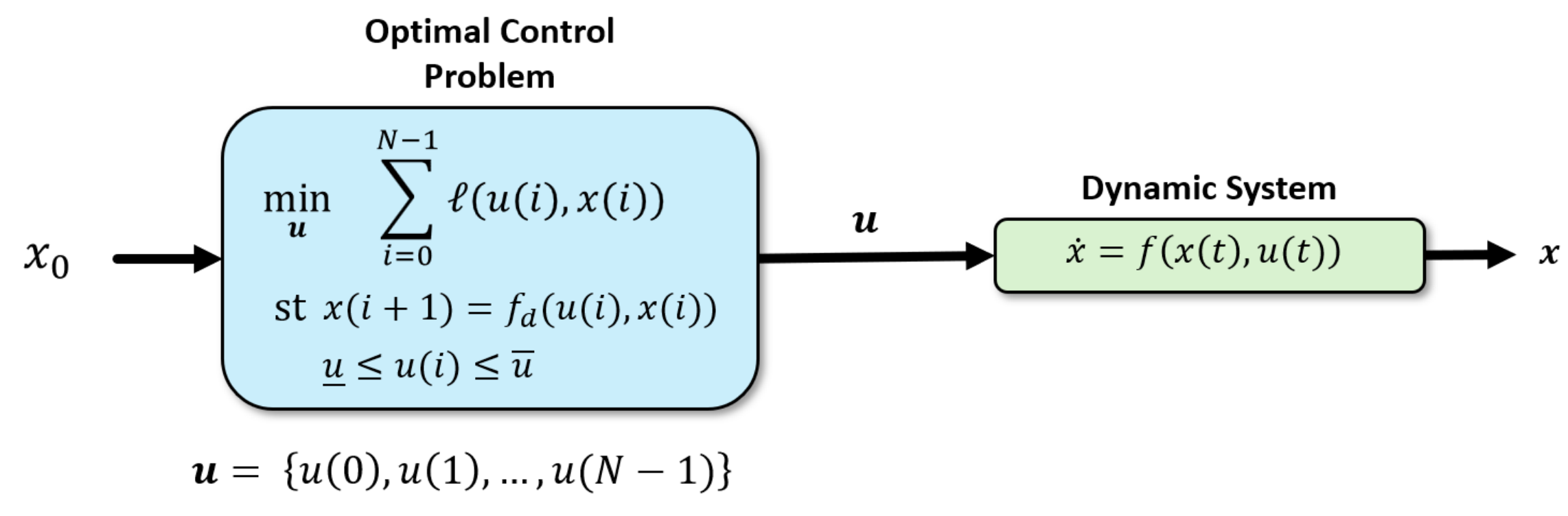}
\caption{Open loop optimal control strategies compute the control sequence~\boldmath$u$ offline prior to the mission by solving an Optimal Control Problem which takes into account the initial state~$x_0$, system dynamics~$f_d$, and other constraints such as input limits~$\bar{u},\underline{u}$. The control sequence is applied to the Dynamic System where~$u(t) = u(i)$ for all~$t \in [i\Delta,i\Delta+\Delta]$, for all~$i=0,\dots,N-1$, and~$\Delta>0$ is the sampling time which results in the state sequence~$\pmb{x}$. Open loop control strategies may not be able to account for perturbations, such as atmospheric drag in the ARPOD problem, since the Optimal Control Problem is solved once before the mission and assumes perfect state information.}
\label{fig:openLoop}
\end{figure}

The control strategy implemented to drive the deputy spacecraft to the docking configuration should (1) achieve the docking configuration, i.e.,~$z \to z_d$, (2) provide robustness to perturbations, e.g., atmospheric drag, J2 perturbations, and solar pressure, (3) account for system constraints, i.e. actuator limits and limited onboard processing capabilities, and (4) consider optimality of performance.
Many different control strategies have been proposed for the ARPOD problem including both closed loop and open loop strategies.
Open loop control strategies include optimal control strategies which solve an optimization problem accounting for system constraints before the execution of the mission and apply the resulting control law without feedback from the system as shown in Figure~\ref{fig:openLoop}.
These control strategies are computationally inexpensive onboard since they are solved offline, but may fail to account for perturbations. Thus, the deputy may not achieve the docking configuration subject to perturbations
which we make specific in Section~\ref{sec:openVClosed}.

\begin{figure}[H]
\centering
\includegraphics[width=0.6\textwidth]{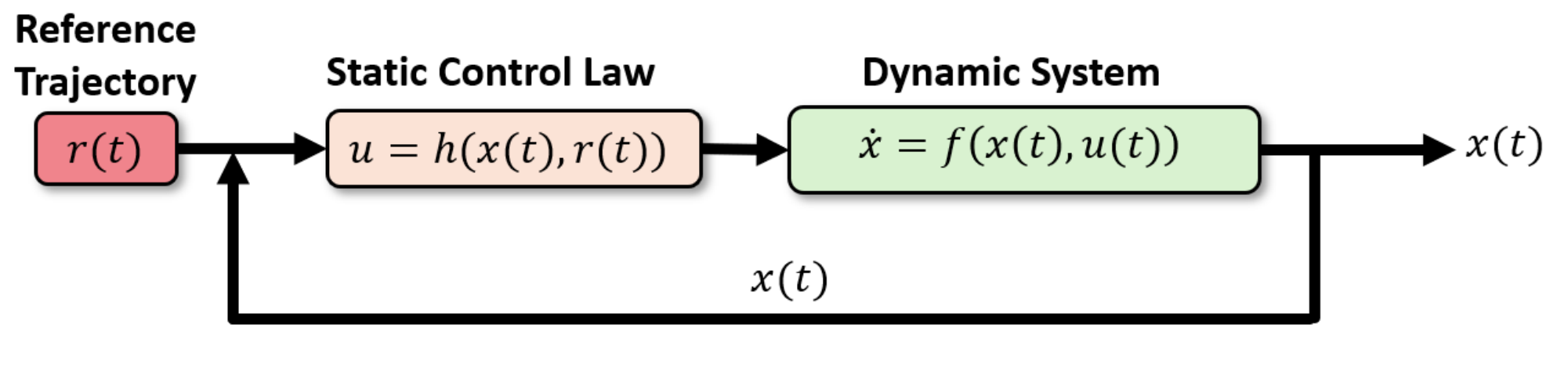}
\caption{Closed loop feedback control laws utilize state feedback from the Dynamic System to drive the state of the system to a desired state, for example the docking configuration in the ARPOD Problem. In this figure, the state~$x(t)$ is fed back to the Static Control Law~$u$ along with the Reference Trajectory~$r(t)$ to compute the next control input. Typically, the static control law is designed to ensure stability of the system and does not consider a performance index for optimality or actuator limits. }
\label{fig:closedLoop}
\end{figure}

Closed loop strategies provide robustness to perturbations by interconnecting the control law within a feedback control loop.
These strategies include static nonlinear control laws with user-defined gains as depicted in Figure~\ref{fig:closedLoop}. 
Typically, static control laws do not take into account any performance index for optimality or other constraints such as actuator limits and rely on a reference trajectory which is either generated offline or computed onboard.
Thus, \emph{a priori} designed control laws may not be suitable for ARPOD missions when a reference trajectory is unavailable and actuator limits must be considered.

Therefore, MPC has been proposed for the ARPOD problem to account for nonlinear dynamics, provide robustness to perturbations, impose actuator constraints, and consider optimality. However, \emph{Conventional} MPC can be computationally expensive and unable to be implemented onboard due to computational time constraints.
In the next section we introduce \emph{Conventional} MPC and show how the \emph{Time-Constrained} MPC strategy we implement differs to account for computational time constraints.

\subsection{Conventional Model Predictive Control} \label{sec:MPC}

Consider the discrete-time dynamics
    $x(k+1) = f_d(x(k),u(k))$,
where~$x(k) \in \R^n$,~$u(k) \in \R^m$, and~$f_d: \R^n \times \R^m \rightarrow \R^n$.
The goal of conventional MPC is to solve an optimal control problem at each timestep~$k$ over a finite prediction horizon of length~$N\in\mathbb{N}$, using the current state as the initial state~$x_0$. This process generates
the control sequence~$\bm{u}^*(k) = \{u^*(k),u^*(k+1),\dots,u^*(N-1)\}$
and the state 
sequence~$\bm{x}^*(k) = \{x^*(k),x^*(k+1),\dots,x^*(N)\}$. 
Then, the first input in this sequence,~$u^*(k)$, is applied to the system where~$u(t) = u^*(k)$ for all~$ t \in [k\Delta ,k \Delta + \Delta]$ and~$\Delta > 0$ is the sampling time. This process is repeated until the end of the time horizon is reached. 
Formally, the conventional MPC strategy that is solved at each time~$k$ is given
next. 
\begin{figure}[H]
\centering
\includegraphics[width=0.6\textwidth]{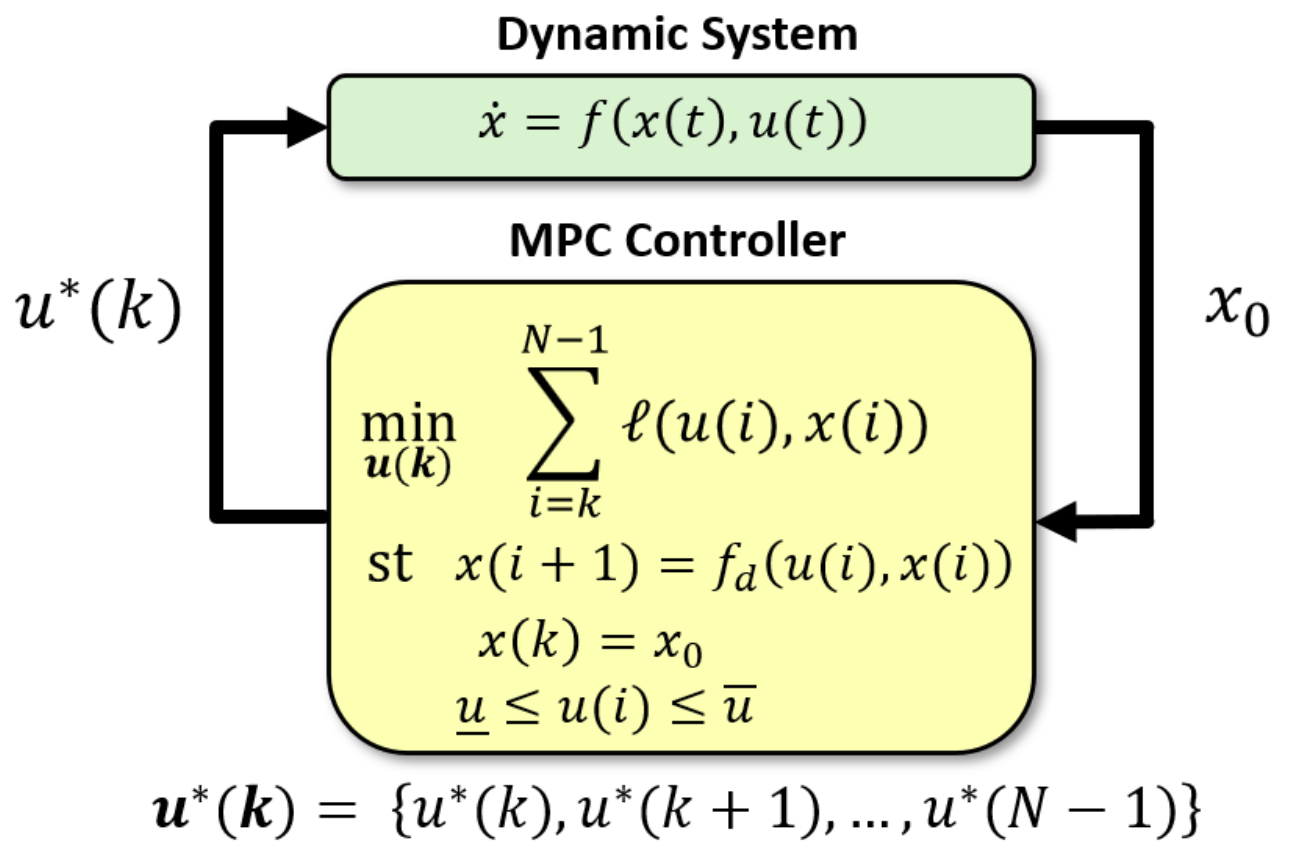}
\caption{At each time step~$k$ the state of the Dynamic System~$x_0$ is provided to the MPC Controller. Then, the MPC controller computes the optimal control sequence~\boldmath$u^*(k)$. Then, the first element of the sequence is applied to the Dynamic System where~$u(t) = u^*(k)$ for all~$ t \in [k\Delta ,k \Delta + \Delta]$ and~$\Delta > 0$ is the sampling time. This process is repeated until the prediction horizon~$N$ is met. Conventional MPC can account for nonlinear dynamics, input limits, and perturbations since the optimal control problem is solved at each time step~$k$. However, it can be computationally expensive and not able to be solved at a fast enough timescale to be implemented online. Thus, computational time constraints should be addressed in the proposed control strategy for the ARPOD problem.
}
\label{fig:mpc}
\end{figure}
In Figure~\ref{fig:mpc},~$\ell : \R^n \times \R^m \rightarrow \reals$
is the cost functional,~$\underline{u}, \bar{u} \in \R^m$ are lower and upper input limits, respectively, and~$\bm{u(k)}=\{u(k),\dots,u(N-1) \}$.
\emph{Conventional} MPC is able to (1) achieve the docking configuration, i.e.,~$z(k) \to z_d$ (which we show in Section~\ref{sec:results}), (2) provide robustness to perturbations (3) account for actuator limits, and (4) consider optimality of performance. However, it may fail account for computational time constraints due to computations performed onboard the deputy processor. 
Specifically,~$\bm{u}^*(k)$ is typically computed using an iterative algorithm which computes a number of iterations until it reaches a stopping condition. Each iterate takes time to compute and it is typically not known beforehand how many iterations will need to be computed to reach the stopping condition. 
Moreover, computing~$\bm{u}^*(k)$ can be computationally burdensome to solve due to nonconvex constraints and suffer from the curse of dimensionality. 
Therefore, we propose to implement \emph{Time-Constrained} MPC which explicitly accounts for computational time constraints by limiting the the number of iterations the underlying optimization algorithm is allowed to complete at each time step~$k$ which we present next.

\subsection{Time-Constrained MPC for ARPOD} \label{sec:tcMPC}
In conventional MPC, the optimization problem is iteratively solved to completion at each time~$k$ by reaching a stopping condition based on the optimality of the solution.
We use~$j_k \in \mathbb{N}$ to denote the number
of iterations that an optimization algorithm must complete to reach
a stopping condition in the computation of~$\bm{u}^*(k)$.
In computationally constrained settings, it cannot be guaranteed that there
is time to execute all~$j_k$ desired iterations because a system input
may be needed before those computations are completed. 
Therefore, we employ the following time-constrained MPC problem 
that has an explicit constraint on~$j_k$. 
\begin{problem}[Time-Constrained MPC for ARPOD] \label{prob:tcMPC}
\begin{small}
\begin{equation} 
    \begin{split}
        \underset{\bm{u}(k)}{\textnormal{minimize}} & \ \sum^{N-1}_{i=k} \ell (x(i),u(i)) \\
        \textnormal{subject to} &\ x(k) = x(0) \\
                                &\  x(i+1) = g_d(x(i),u(i)), \\
                                &\ \underbar{u} \leq u(i) \leq \bar{u}, \\
                                & \ j_k \leq j_\max,
    \end{split}
\end{equation}
\end{small}
where~$j_k$ is the number of iterations computed at time~$k$,~$j_\max$ is the maximum allowable iterations for all~$k$, and~$g_d:\R^{13} \times \R^6 \rightarrow \R^{13}$ are the discretized versions 
of~\eqref{eq:CWdynamics_translational},~\eqref{eq:ugly_ass_equation},~\eqref{eq:scalarQuatDynamics}, and~\eqref{eq:vectorQuatDynamics}. 
\end{problem}

\begin{figure}[H]
\centering
\includegraphics[width=0.5\textwidth]{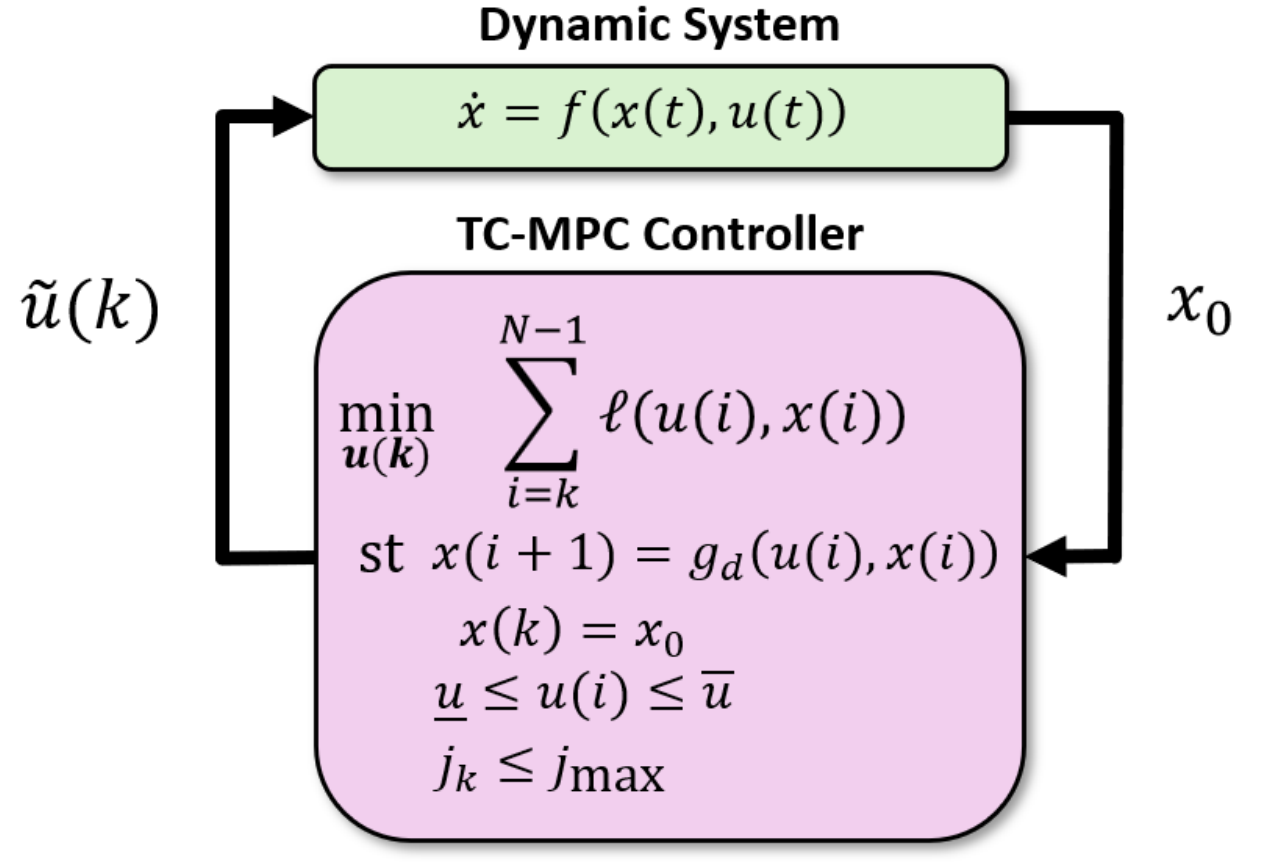}
\caption{At each time step~$k$ the state of the Dynamic System~$x_0$ is provided to the TC-MPC Controller. Then, the TC-MPC controller computes a number of algorithmic iterations~$j_k$ to compute the possibly sub-optimal control sequence~\boldmath$\tilde{u}(k)$. Then, the first element of the sequence is applied to the Dynamic System  where~$u(t) = \tilde{u}(k)$ for all~$ t \in [k\Delta ,k \Delta + \Delta]$ and~$\Delta > 0$ is the sampling time.. This process is repeated until the prediction horizon or a stopping condition is met.
}
\label{fig:tcmpc}
\end{figure}

The constraint~$j_k \leq j_{max}$ models scenarios with limited onboard computational speed in which there is only enough time to complete at most~$j_\max$ iterations. 
A solution to Problem~\ref{prob:tcMPC} at time~$k$ may
result in sub-optimal input and state sequences, i.e.,
\begin{equation}
    \begin{split}
        \tilde{\bm{u}}(k) &= \{\tilde{u}(k),\tilde{u}(k+1),\dots,\tilde{u}(N-1)\} \\
        \tilde{\bm{x}}(k) &= \{\tilde{x}(k),\tilde{x}(k+1),\dots,\tilde{x}(N)\}, \\
    \end{split}
\end{equation}
respectively. Then, we apply the first input of the resulting input 
sequence, namely~$\tilde{\bm{u}}(k)$, and repeat this process until the 
end of the time horizon is reached. 
This setup is different from conventional MPC in that we apply a potentially sub-optimal input due to the iteration constraint~$j_k \leq j_{max}$ in Problem~\ref{prob:tcMPC}.
Figure~\ref{fig:tcmpc} shows the block diagram for our proposed time-constrained MPC strategy.

\subsection{Open Loop Optimal Control vs. MPC} \label{sec:openVClosed}
\begin{figure}[H]
\centering
\includegraphics[width=0.6\textwidth]{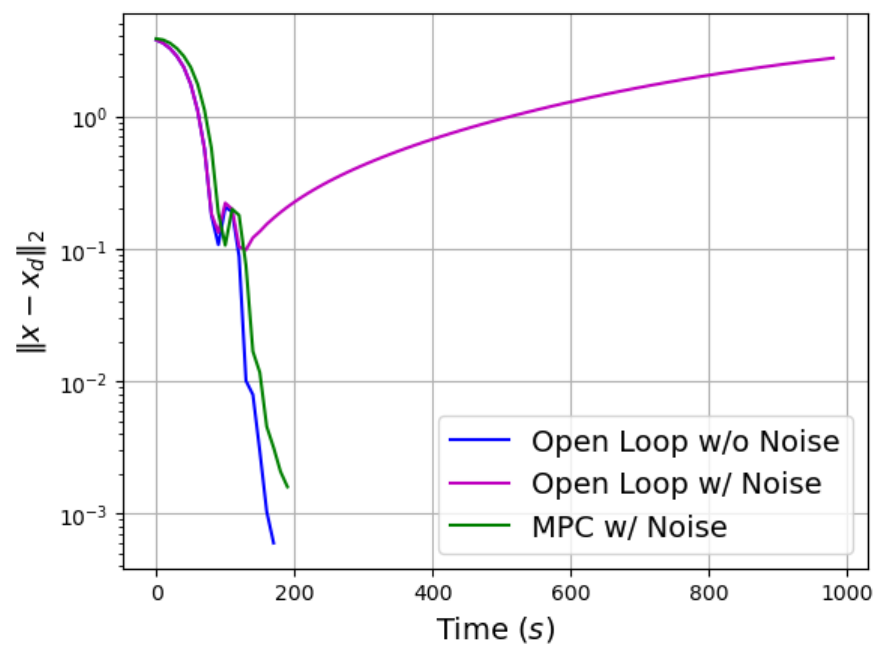}
\caption{ This plot displays the deputy's state error that results from applying an open loop optimal control sequence~$\bm{u}_{OL}$ and closed loop MPC strategy~$\bm{u}_{MPC}$ where the dynamics are subject to perturbations. The open loop optimal control sequence~~$\bm{u}_{OL}$ is unable to drive the deputy state to the docking state since it is computed \emph{a priori} without knowledge of perturbations. The MPC sequence~$\bm{u}_{MPC}$ is able to drive the deputy state to the docking state subject to perturbations because it utilizes state feedback to update~$\bm{u}_{MPC}$ during the mission.
}
\label{fig:openVClosed}
\end{figure}

To demonstrate the robustness of MPC to perturbations as compared to an open loop optimal control strategy we simulate both strategies subject to perturbations. First, for the open loop solution, we solved the following problem
\begin{equation} \label{eq:open}
    \begin{split}
        \underset{\bm{u}}{\textnormal{minimize}} & \ \sum^{N-1}_{i=0} (x(i)-x_d)^\top Q (x(i)-x_d) + u(i)^\top R u(i) \\
        \textnormal{subject to} &\ x(0) = x_0, \\
                                &\  x(i+1) = g_d(x(i),u(i)), \\
                                &\ \underline{u} \leq u(i) \leq \bar{u},
    \end{split}
\end{equation}
where~$N$,~$Q \in \R^{13\times 13}$, and~$R \in \R^{6\times 6}$ are defined in Table~\ref{tb:parameters},~$\underline{u} \coloneqq (-10^{-2} \times \ones^\top_3, \ -10^{-4} \times \ones^\top_3)^\top$,~$\bar{u} \coloneqq -\underline{u}$,~$x_0=[1.5, -1.77, 3,10^{-3},3.4\times 10^{-3}, 0, 0.772,0.463,0.309,0.309,-2.15 \times 10^{-4},10^{-3},-4.6 \times 10^{-3}]$, and~$g_d$ is generated by the 4th order Runge-Kutta method. 
The solution to~\eqref{eq:open} yields an optimal control sequence we denote by~$\bm{u}_{OL} = \{u_{OL}(0),\dots,u_{OL}(N-1) \}$ where~$u_{OL}(i) \in \R^6$ for all~$i \in [0,N-1]$. 
To simulate the open loop control strategy subject to perturbations, at each time~$i$, we add a perturbation~$\omega(i) \sim \mathcal{U}_{13}(0,10^{-4})$ to the dynamics. 
Specifically, at each time~$i$, we apply~$u_{OL}(i)$ and~$\omega(i)$ to the dynamic system which results in~$x(i+1) = g_d(x(i), u_{OL}(i)) + \omega(i)$. 
Figure~\ref{fig:openVClosed} shows the state error that results from applying~$\bm{u}_{OL}$ with and without perturbations.
The open loop control sequence~$\bm{u}_{OL}$ is able to drive the deputy to the docking configuration when no perturbations are added. When perturbations are included~$\bm{u}_{OL}$ does not drive~$x \to x_d$ because the control law does not incorporate state feedback from the system and is unable to adapt.

\begin{figure}[H]
  \centering
  \begin{tabular}{cc}
    \begin{subfigure}{0.45\textwidth}
      \centering
    \includegraphics[width=\linewidth]{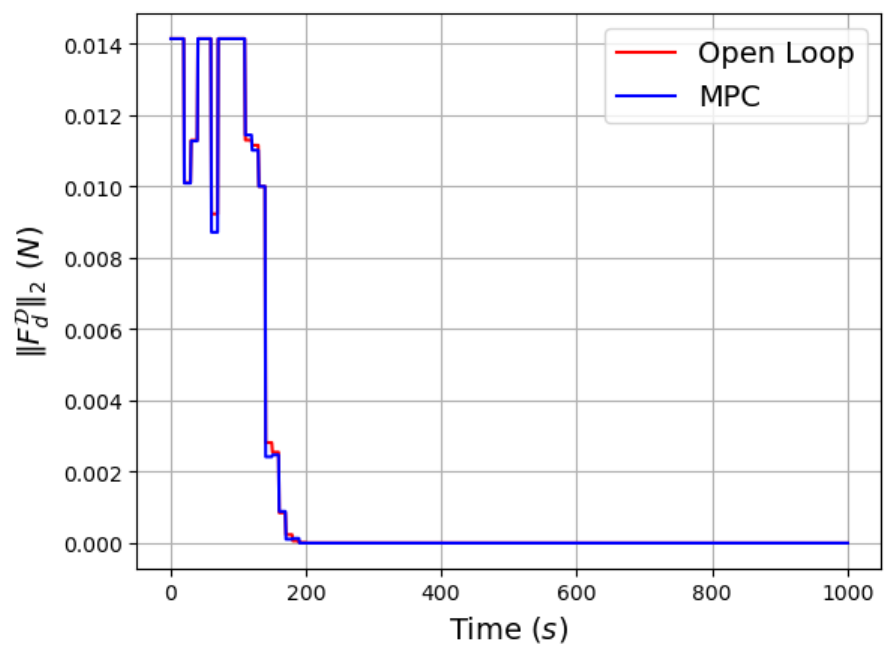}
      \caption{Open Loop Optimal Control and Closed Loop MPC Thrust Magnitude}
      \label{fig:openVClosedThrust}
    \end{subfigure} &
    \begin{subfigure}{0.45\textwidth}
      \centering
    \includegraphics[width=\linewidth]{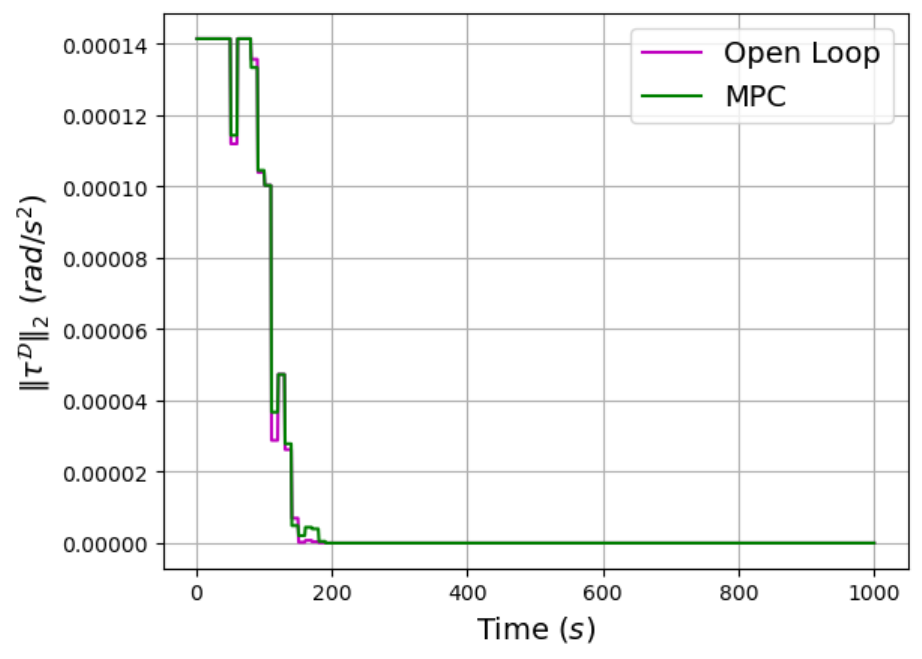}
      \caption{Open Loop Optimal Control and Closed Loop MPC Torque Magnitude}
      \label{fig:openVClosedTorque}
    \end{subfigure} \\
  \end{tabular}
  \caption{These plots display the magnitudes of the deputy thrust and torques over the ARPOD mission for the initial condition specified in this section computed using Open Loop Optimal Control and MPC subject to perturbations. The MPC control inputs differ from the Open Loop Optimal control inputs because the MPC controller computes new control sequences over the course of the mission using feedback from the dynamic system to account for perturbations. }
  \label{fig:openVClosedInputs}
\end{figure}

Next, we simulated conventional MPC as shown in Figure~\ref{fig:mpc} subject to perturbations to demonstrate its robustness.
At each time~$k$, the optimal control sequence is solved for which we denote as~$\bm{u}_{MPC}(k) = \{ u_{MPC}(k), \dots, u_{MPC}(N-1) \}$. Then we apply~$u_{MPC}(k)$ and~$\omega(k)$ to the dynamic system which results in~$x(k+1) = g_d(x(k), u_{OL}(k)) + \omega(k)$. The perturbed state is then fed back to the MPC controller by setting~$x_0 = g_d(x(k), u_{OL}(k)) + \omega(k)$, the MPC controller computes~$\bm{u}_{MPC}(k+1)$, and this process is repeated. Figure~\ref{fig:openVClosed} shows the resulting state error from implementing MPC subject to the same perturbations as the open loop simulation.
The MPC strategy drives the deputy's state to the docking state, i.e.,~$x \to x_d$, subject to perturbations by utilizing state feedback and recomputing its control sequence~$\bm{u}_{MPC}(k)$ at each~$k$ over the course of the ARPOD mission.
This enables MPC to mitigate the effect of perturbations which open loop optimal control strategies are not able to mitigate.
The magnitudes of the deputy thrust and torque profiles are shown in Figure~\ref{fig:openVClosedInputs}.


\section{Hardware-in-the-Loop Processing} \label{sec:simulation}
\subsection{Hardware} \label{sec:hardware}
\begin{figure}
  \centering
  \begin{tabular}{cc}
    \begin{subfigure}{0.3\textwidth}
      \centering
    \includegraphics[width=\linewidth]{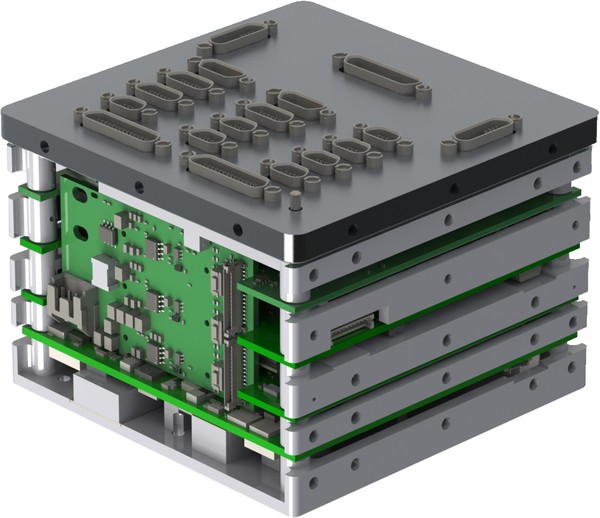}
      \caption{CAD model of iX10-101.}
      \label{fig:Unibap}
    \end{subfigure} &
    \begin{subfigure}{0.6\textwidth}
      \centering
    \includegraphics[width=\linewidth]{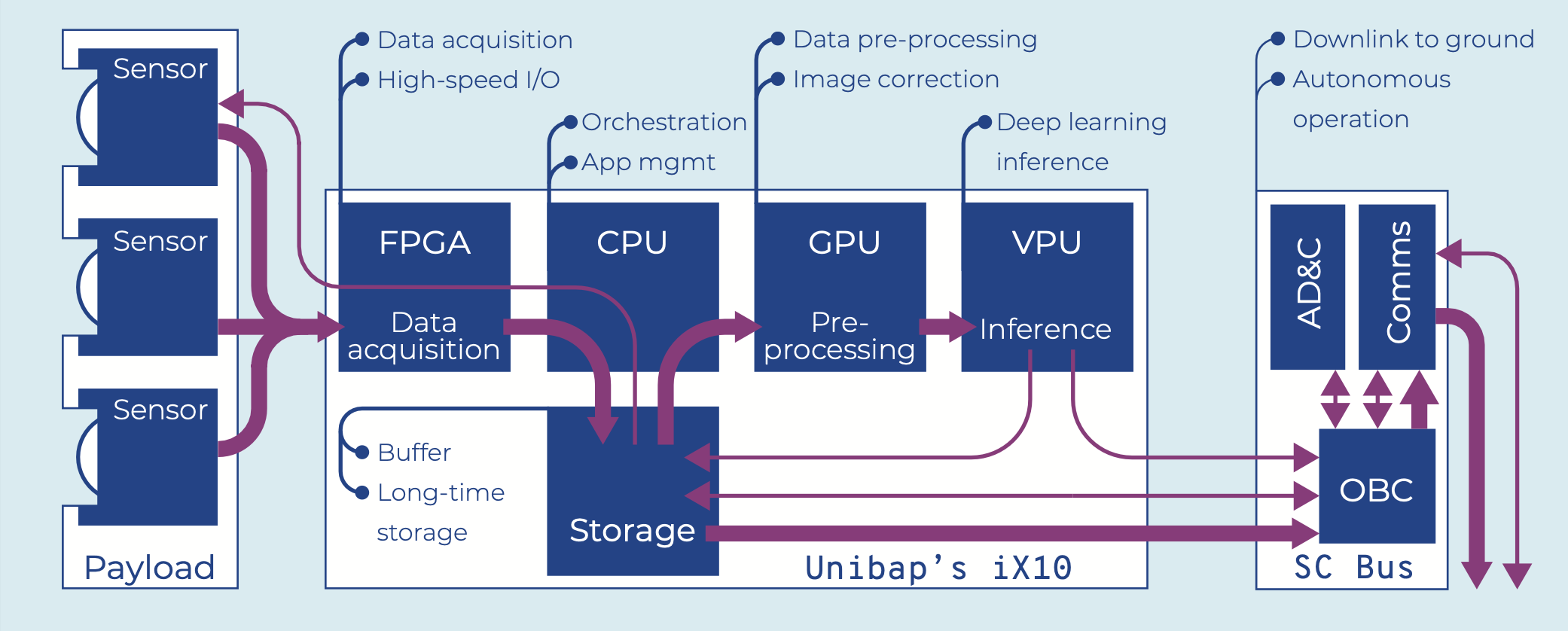}
      \caption{Depiction of data flow into and out out of the interfaces of the iX10.}
      \label{fig:Dataflow}
    \end{subfigure} \\
  \end{tabular}
  \caption{The above are taken from the Unibap iX10-101 datasheet \cite{unibap_ds}. }
  \label{fig:hardware}
\end{figure}

Benchmarking was performed at the Spacecraft Performance Analytics and Computing Environment Research (SPACER) laboratory housed within the Space Vehicles directorate at Kirtland Air Force Base. The simulations described below in Section~\ref{sec:results} were implemented onboard the Unibap iX10-101, a radiation-tolerant\footnote{It is the authors' understanding that the amount of radiation tolerance for this board is applicable in most LEO-MEO orbits but not GEO orbits.} board which houses an AMD v1605b processor. With regard to modeling and simulation setup, C++ code modified for Unibap OS (Linux-based) was run on the CPU and the GPU was not utilized in the present study. It should be noted that all data incoming through the payload subsystem (Figure~\ref{fig:Dataflow}) was emulated onboard at each state acquisition step at each timestep $k$. Similarly, commanded actuation to the AD\&C subsystem (Figure~\ref{fig:Dataflow}) was emulated through the handoff of the computed control $u_{MPC}(k)$.

\subsection{Time-Constrained Optimization Results} \label{sec:results}
In this section we empirically validate the use of time-constrained MPC in Figure~\ref{fig:tcmpc} for the dynamics which model the 6 DOF ARPOD problem in~\eqref{eq:CWdynamics_translational},~\eqref{eq:ugly_ass_equation},~\eqref{eq:scalarQuatDynamics}, and~\eqref{eq:vectorQuatDynamics} by running Monte Carlo experiments on space grade hardware described in Section~\ref{sec:hardware}.
The problem we solve in this section is given next.
\begin{problem}[Time-Constrained MPC for ARPOD Experiments] \label{prob:arpodMPC}
\begin{equation} 
    \begin{split}
        \underset{\bm{u}(k)}{\textnormal{minimize}} &\  \sum^{N-1}_{i=k} (x(i)-x_d)^\top Q (x(i)-x_d) + u(i)^\top R u(i) \\
        \textnormal{subject to} &\ x(k) = x_0 \\
                                &\  x(i+1) = g_d(x(i),u(i)) \\
                                &\ \underbar{u} \leq u(i) \leq \bar{u}, \\
                                & \ j_k \leq j_\max,
    \end{split}
\end{equation}
where~$\bar{u} \in \R^6$,~$Q \in \R^{13\times 13}$ and~$R \in \R^{6\times 6}$ are defined in Table~\ref{tb:parameters},~$\underbar{u} \coloneqq -\bar{u}$, and~$g_d:\R^{13} \times \R^6 \rightarrow \R^{13}$ are the discretized versions 
of~\eqref{eq:CWdynamics_translational},~\eqref{eq:ugly_ass_equation},~\eqref{eq:scalarQuatDynamics}, and~\eqref{eq:vectorQuatDynamics} generated by the forward Euler method.
\end{problem}


We solve Problem~\ref{prob:arpodMPC} numerically for 200 unique initial conditions 
for~$\jmax \in \{ 1,2,3,4,5,6,7,8,9,10,50,100, \text{Optimal} \}$, where~${\jmax=\text{Optimal}}$ excludes the maximum allowable iteration constraint in Problem~\ref{prob:arpodMPC}.
We generate the initial conditions such that the initial state~$x_{\text{init}} \in \bar{x} \times -\bar{x}$ where~$\bar{x} = [1.5 \times \ones^\top_3 \text{ km}, 0.001 \times \ones^\top_3 \text{ km/s}, \ones^\top_4, 0.002 \times \ones^\top_3 \text{ rad/s} ]^\top$.
We consider the parameters in Table~\ref{tb:parameters} where~$n$ is the chief mean motion,~$m_d$ is the deputy mass,~$J_d$ is the deputy moment of inertia matrix,~$t_s$ is the sampling time,~$N$ is the prediction horizon, and~$Q$ and~$R$ are cost matrices.

\begin{table}[hb]
\begin{center}
\caption{Problem Parameters}\label{tb:parameters}
\begin{tabular}{c c c c c} 
 \toprule
 \toprule
 Parameter Name & Parameter & Value & Units  \\  
 \toprule
Mean motion&$n$ & -0.0011 & rad/s \\
Deputy Mass &$m_d$ & 12 & kg \\ 
Deputy Moment of Inertia &$J_d$ & $\textnormal{diag}\{[0.2734, 0.2734, 0.3125]\}$ & kg$\cdot \text{m}^2$ \\
Sampling Time&$t_s$ & 10 & s \\ 
Prediction Horizon &$N$ & 100 & - \\ 
State Cost Matrix&$Q$ & $\textnormal{diag}\{[10^{5}\times \ones^\top_3, 10^{2}\times \ones^\top_3, 10^{6}\times \ones^\top_4,
10^{7}\times \ones^\top_3]\}$ & - \\ 
Input Cost Matrix &$R$ & $\textnormal{diag}\{[10^{5}\times \ones^\top_3, 10^{10}\times \ones^\top_3]\}$ & - \\ 
Input Limits & \emph{$\bar{u} $} & $(10^{-2} \times \ones^\top_3, \ 10^{-4} \times \ones^\top_3)^\top$ & N, $\text{rad/s}^2$ \\
 \toprule
 \toprule
\end{tabular}
\end{center}
\end{table}
These choices of~$Q$ and~$R$ penalize the error in the attitudinal states, i.e.,~$\delta q_{\D}^{\mathcal{O}}, \delta \omega_{\D}^{\mathcal{O}}, \tau^\D$, more than the translational states, i.e.~$\delta r^{\mathcal{O}}, \delta \dot{r}^{\mathcal{O}}, F^\D_d$.
We designed~$Q$ and~$R$ in this way to address the coupling of the attitudinal and translational dynamics, however other choices of~$Q$ and~$R$ can be selected by the user to prioritize other objectives.
We formulated Problem~\ref{prob:arpodMPC} in C++ using the CasADi symbolic framework~\cite{andersson2019casadi} and solved it using IPOPT~\cite{wachter2006implementation}.\footnote{Our code can be found at https://github.com/gbehrendt/Satellite.}
In the minimization of Problem~\ref{prob:arpodMPC} and at each~$k$, we designed two stopping conditions which terminate the minimization if met: (1) optimality error of the problem is less than~$10^{-5}$\footnote{See~\cite[Equation 6]{wachter2006implementation} for the definition of ``optimality error''} or (2) the algorithm had completed the maximum allowable number of iterations, i.e.~$j_k = \jmax$, which is how we enforce the computational time constraint.
In simulation, we say the deputy has achieved a successful docking configuration when~$\| z(k) - z_d \|_\infty \leq 10^{-3}$, i.e., each element of~$z(k)$ is within an error ball of size~$10^{-3}$ of the desired value.
Table~\ref{tab:converge} shows the number of trials that achieved a successful docking configuration for each value of~$\jmax$. 
As~$\jmax$ increased more trials achieve the docking configuration for~$\jmax=1,2,3,4,5$ and for~$\jmax \geq 6$ all 200 trials converge due to the optimizer completing more iterations.
\begin{table}
\centering
 \caption{Number of trials that achieved a successful docking configuration}
\begin{tabular}{|c | c c c c c c c c c c c c c|} 
 \hline
 $\jmax$  & 1 & 2 & 3 & 4 & 5 & 6 & 7 & 8 & 9 & 10 & 50 & 100 & Optimal  \\ 
 \hline
 Success & 0 & 112 & 151 & 174 & 194 & 200 & 200 & 200 & 200 & 200 & 200 & 200 & 200 \\ 
 \hline
 Failed & 200 & 88 & 49 & 26 & 6 & 0 & 0 & 0 & 0 & 0 & 0 & 0 & 0 \\ 
 \hline
\end{tabular}
\label{tab:converge}
\end{table}

\begin{figure}[H]
  \centering
  \begin{tabular}{cc}
    \begin{subfigure}{0.45\textwidth}
    \centering
    \includegraphics[width=\linewidth]{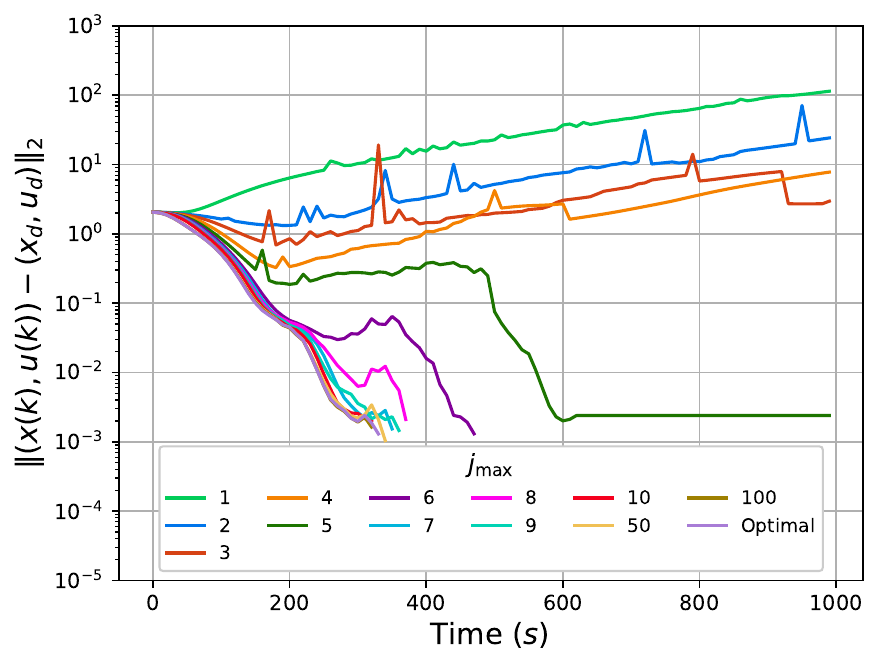}
    \caption{Average Error}
    \label{fig:avgError}
    \end{subfigure} &
    \begin{subfigure}{0.45\textwidth}
          \centering
    \includegraphics[width=\linewidth]{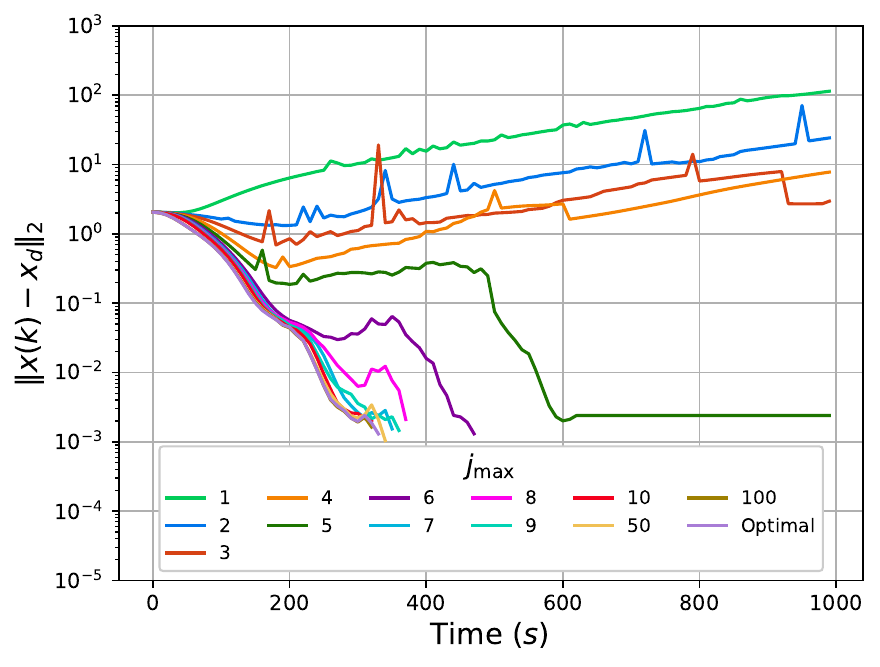}
    \caption{Average State Error}
    \label{fig:avgErrorState}
    \end{subfigure} \\
    \begin{subfigure}{0.45\textwidth}
          \centering
    \includegraphics[width=\linewidth]{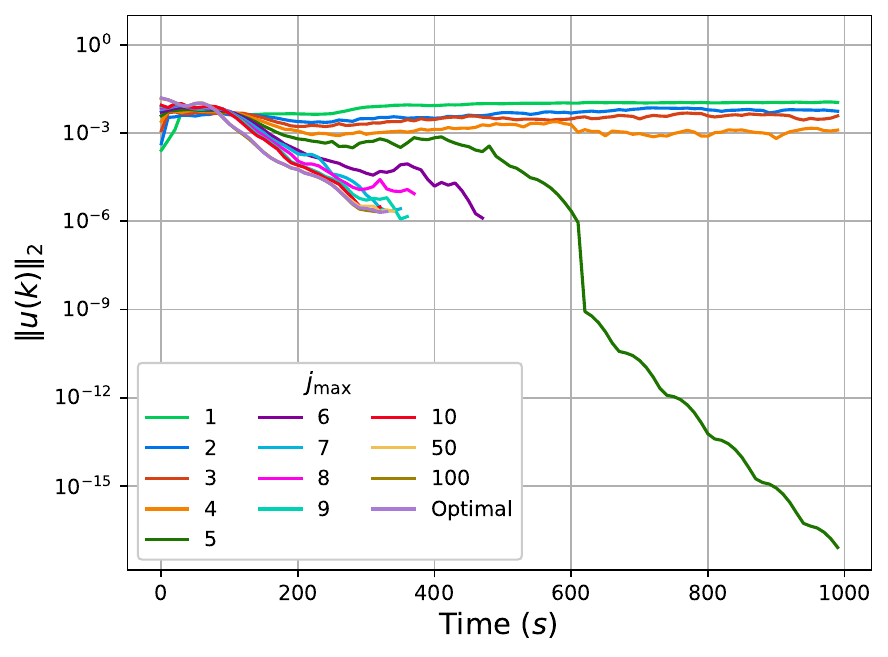}
    \caption{Average Control Error}
    \label{fig:avgErrorControl}
    \end{subfigure} &
    \begin{subfigure}{0.45\textwidth}
          \centering
    \includegraphics[width=\linewidth]{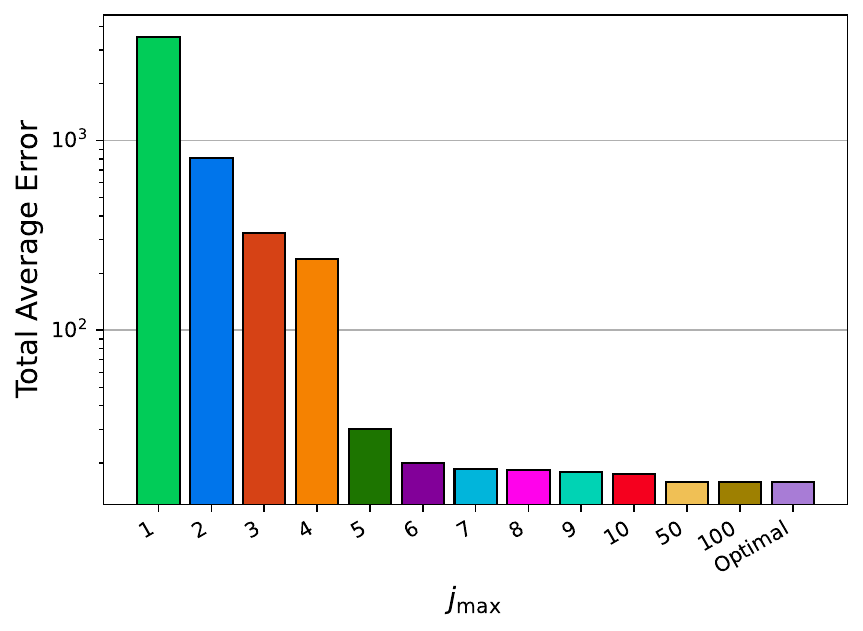}
    \caption{Total Average Error}
    \label{fig:totAvgError}
    \end{subfigure} 
  \end{tabular}
  \caption{These plots display the Average Error, Average State Error, Average Control Error, and Total Average Error. 
  As the number of maximum allowable iterations~$\jmax$ increases the TC MPC strategy incurs less error as it drives the state and control inputs of the deputy satellite to the docking configuration in Figure~\ref{fig:totAvgError}.}
  \label{fig:error}
\end{figure}

Figure~\ref{fig:avgError} shows the average error, i.e.,~$\| z(k) - z_d \|_2$, across all 200 trials of the deputy spacecraft for all simulated values of~$\jmax$. 
Figures~\ref{fig:avgErrorState} and~\ref{fig:avgErrorControl} show the state~$\| x(k) - x_d \|_2$ and control error~$\| u(k) \|_2$, respectively. 
In Figure~\ref{fig:totAvgError}, we take the sum of the average error in Figure~\ref{fig:avgError} to obtain the total average error.

\begin{figure}[H]
        \centering
        \begin{subfigure}[b]{0.45\textwidth}
            \centering
            \includegraphics[width=\textwidth]{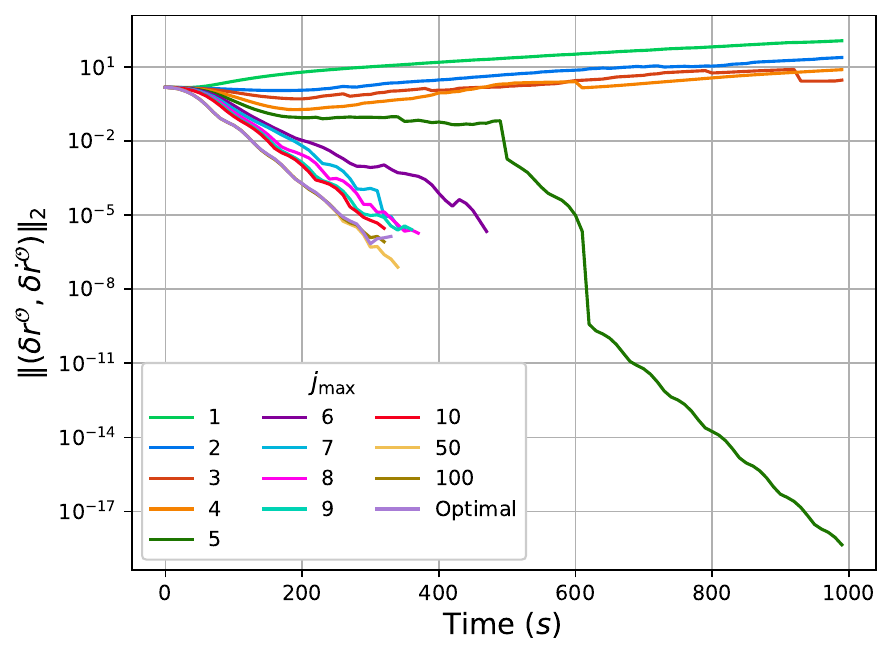}
            \caption[]%
            {{Translational Error}}    
            \label{fig:transError}
        \end{subfigure}
        \hfill
        \begin{subfigure}[b]{0.45\textwidth}   
            \centering 
            \includegraphics[width=\textwidth]{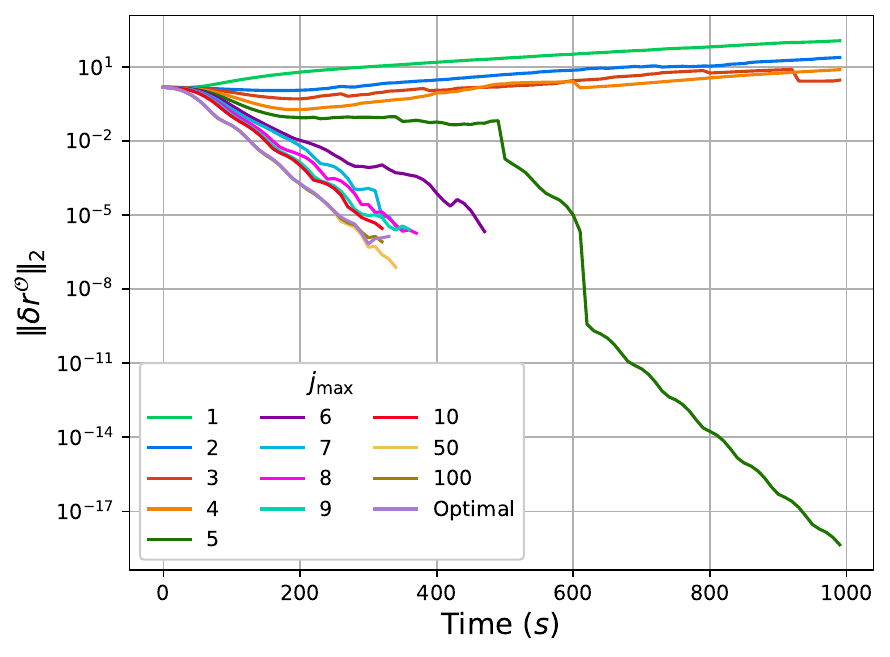}
            \caption[]%
            {{Position,~$\delta r^{\coordframe{O}}$ (km)}}    
            \label{fig:position}
        \end{subfigure}
        \vskip\baselineskip
        \begin{subfigure}[b]{0.45\textwidth}   
            \centering 
            \includegraphics[width=\textwidth]{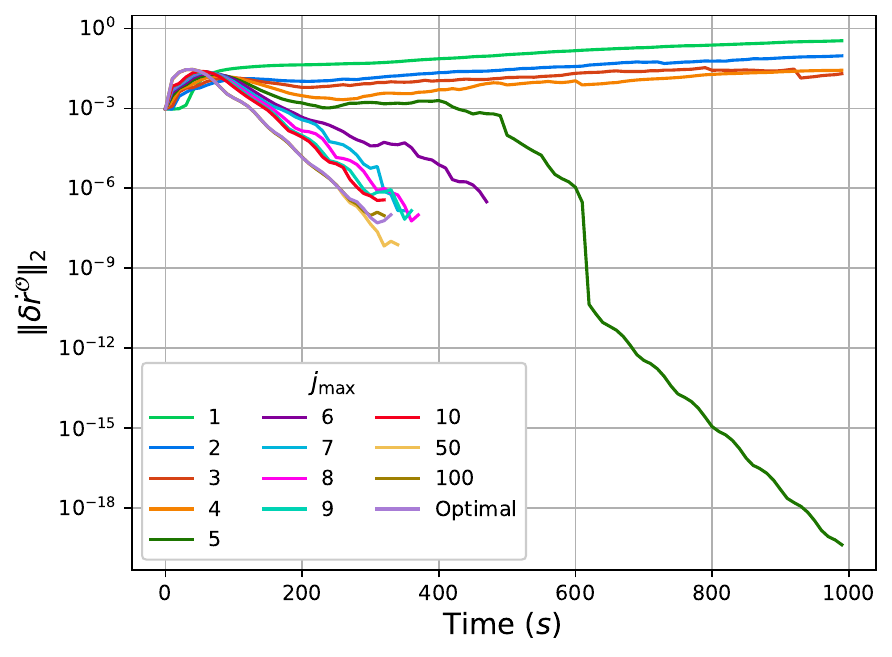}
            \caption[]%
            {{Linear Velocity,~$\delta \dot{r}^{\coordframe{O}}$ (km/s)}}    
            \label{fig:velocity}
        \end{subfigure}
        \hfill
        \begin{subfigure}[b]{0.45\textwidth}  
            \centering 
            \includegraphics[width=\textwidth]{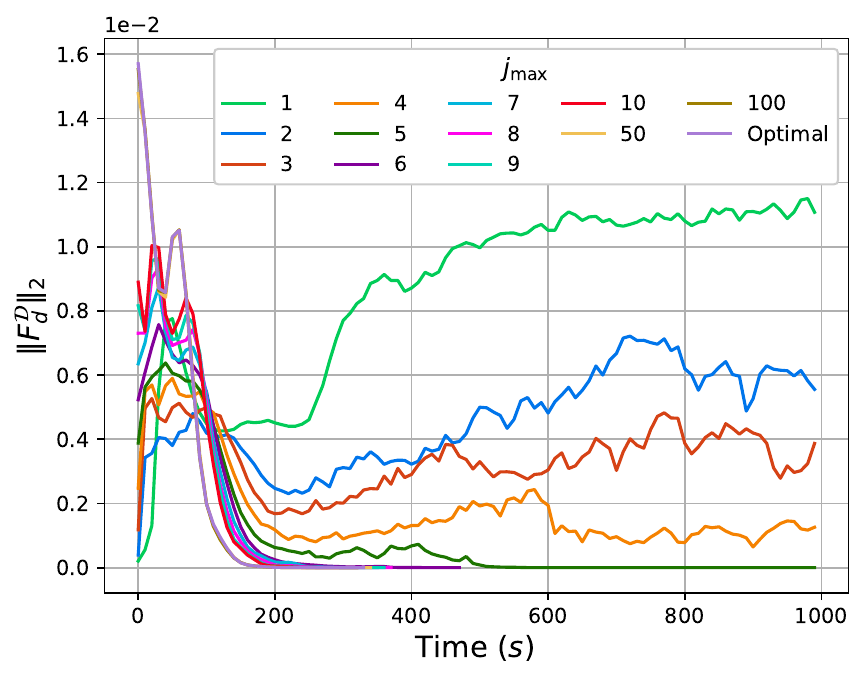}
            \caption[]%
            {{Thrust Inputs,~$F^{\coordframe{D}}_d$ (N)}}    
            \label{fig:thrust}
        \end{subfigure}
        \caption[]
        {These plots display the translational error~\ref{fig:transError}, positional error~\ref{fig:position}, linear velocity error~\ref{fig:velocity}, and deputy thrust profile~\ref{fig:thrust}. As the number of maximum allowable iterations~$\jmax$ increases the TC MPC strategy drives the translational states to the docking configuration in less time.} 
        \label{fig:translation}
\end{figure}

\begin{figure}[H]
        \centering
        \begin{subfigure}[b]{0.45\textwidth}
            \centering
            \includegraphics[width=\textwidth]{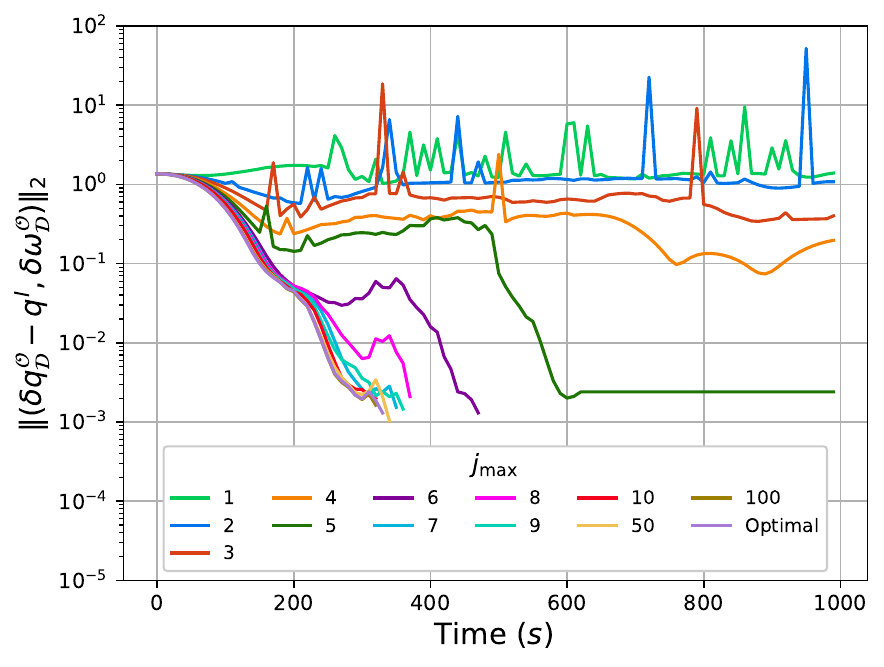}
            \caption[]%
            {{Attitudinal Error}}    
            \label{fig:rotError}
        \end{subfigure}
        \hfill
        \begin{subfigure}[b]{0.45\textwidth}   
            \centering 
            \includegraphics[width=\textwidth]{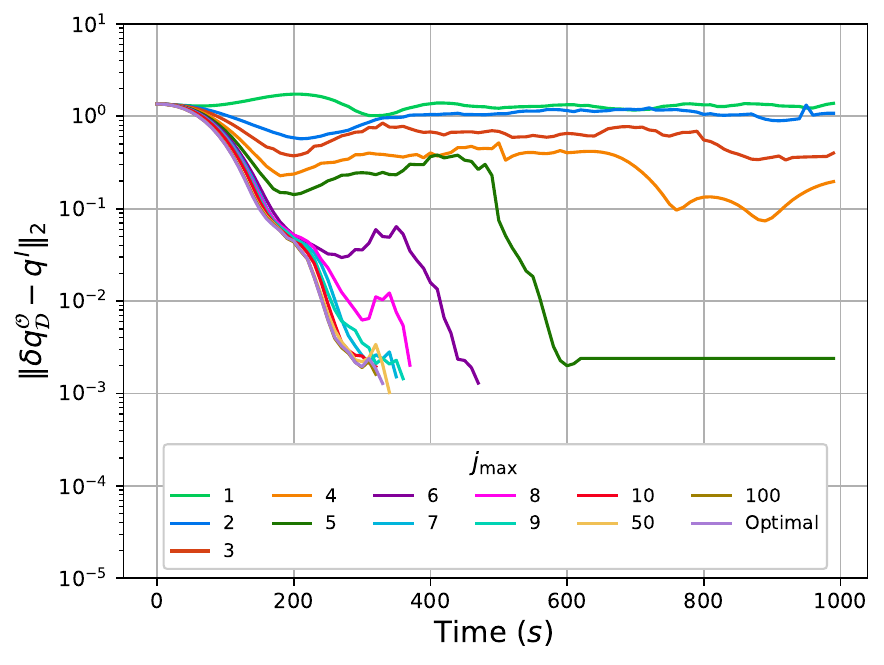}
            \caption[]%
            {{Error Quaternion,~$\delta q^{\coordframe{O}}_{\coordframe{D}}$}}    
            \label{fig:quaternion}
        \end{subfigure}
        \vskip\baselineskip
        \begin{subfigure}[b]{0.45\textwidth}   
            \centering 
            \includegraphics[width=\textwidth]{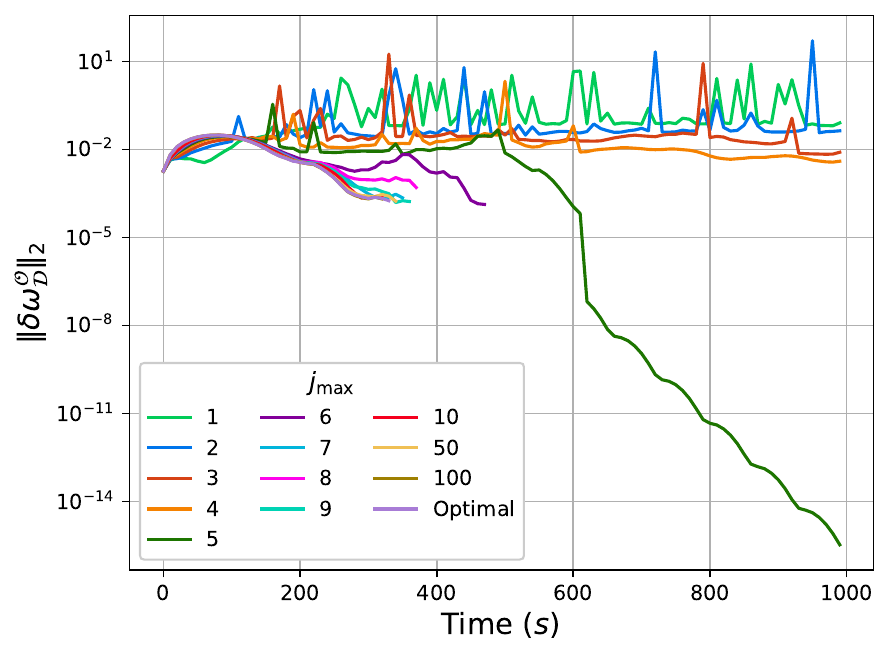}
            \caption[]%
            {{Error Angular Velocity,~$\delta \omega^{\coordframe{O}}_{\coordframe{O}\coordframe{D}} (\text{rad/s})$ }}    
            \label{fig:angularVel}
        \end{subfigure}
        \hfill
        \begin{subfigure}[b]{0.45\textwidth}  
            \centering 
            \includegraphics[width=\textwidth]{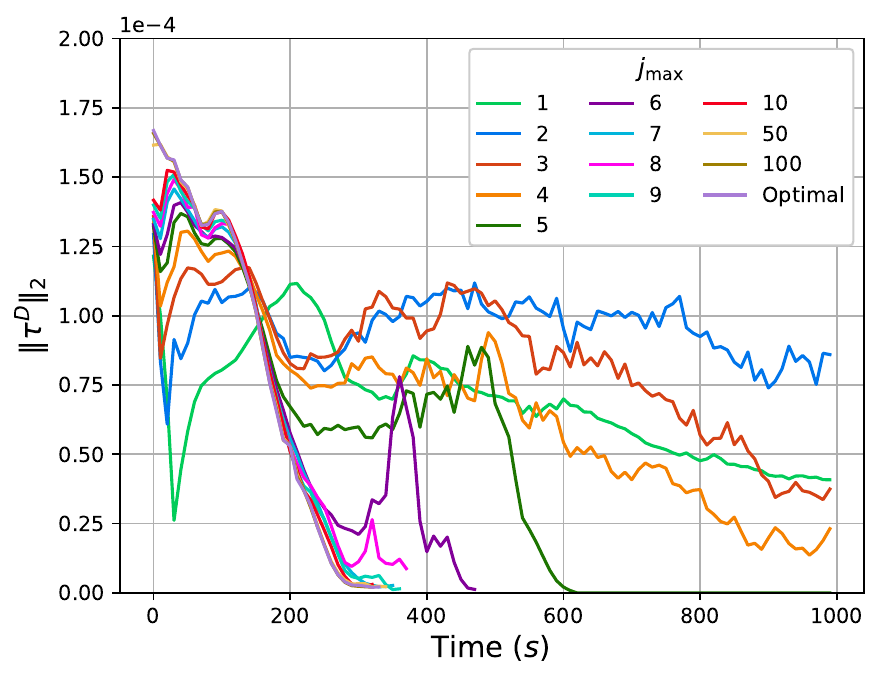}
            \caption[]%
            {{Torque Inputs,~$\tau^{\coordframe{D}} (\text{rad/s}^2$}) }    
            \label{fig:torque}
        \end{subfigure}
        \caption[]
        {These plots display the attitudinal error~\ref{fig:rotError}, quaternion error~\ref{fig:quaternion}, angular velocity error~\ref{fig:angularVel}, and deputy torque profile~\ref{fig:torque}. As the number of maximum allowable iterations~$\jmax$ increases the TC MPC strategy drives the attitudinal states to the docking configuration in less time.} 
        \label{fig:attitude}
\end{figure}

\begin{figure}[H]
  \centering
  \begin{tabular}{cc}
    \begin{subfigure}{0.45\textwidth}
      \centering
    \includegraphics[width=\linewidth]{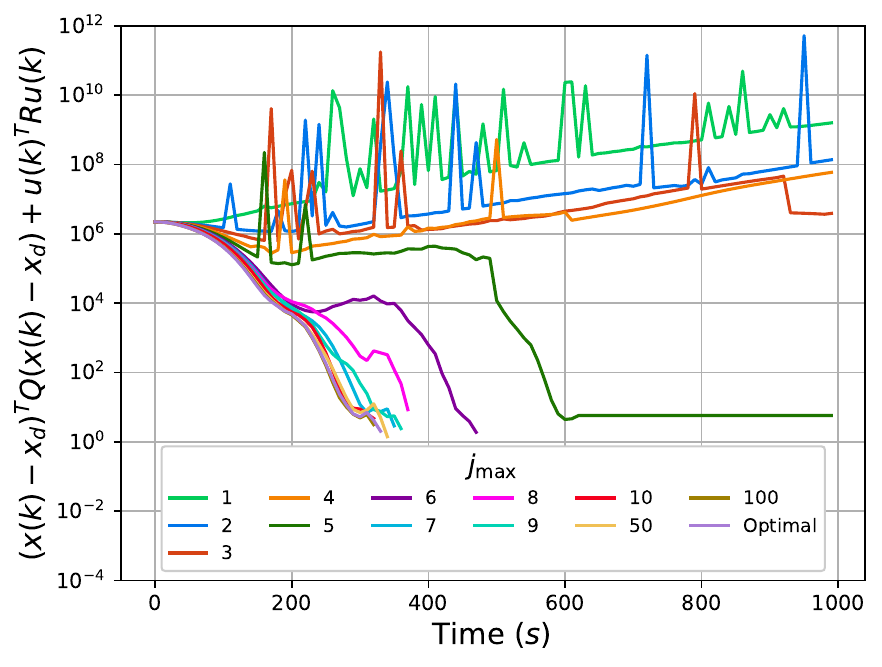}
      \caption{Average Cost}
      \label{fig:avgCost}
    \end{subfigure} &
    \begin{subfigure}{0.45\textwidth}
      \centering
    \includegraphics[width=\linewidth]{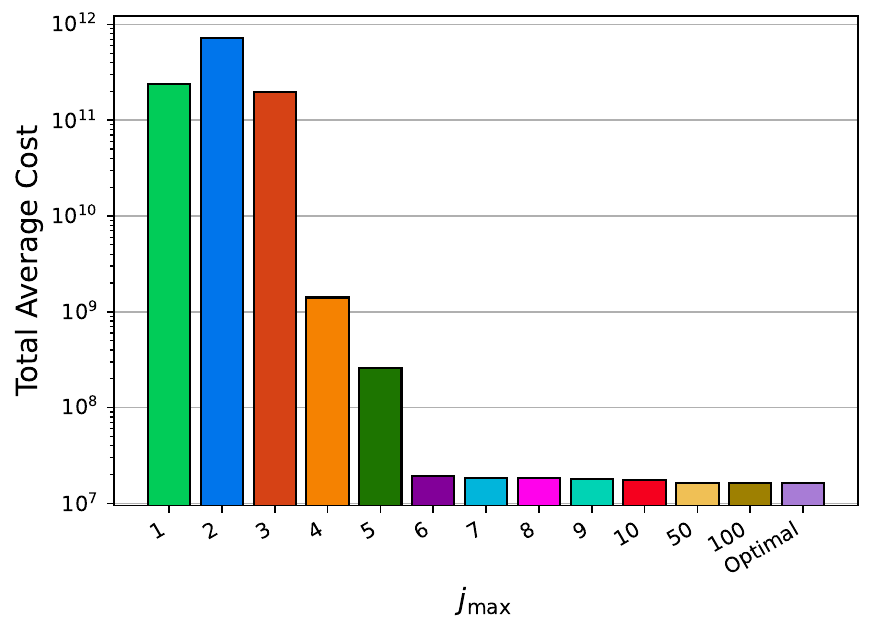}
      \caption{Total Average Cost}
      \label{fig:totAvgCost}
    \end{subfigure} \\
  \end{tabular}
  \caption{These plots display the average cost over time~\ref{fig:avgCost} and total average cost~\ref{fig:totAvgCost}. We observe that as the number of maximum allowable iterations~$\jmax$ increases the deputy incurs less cost when solving Problem~\ref{prob:arpodMPC}.}
  \label{fig:cost}
\end{figure}

Moreover, the average translational~$\left\| \left( \delta r^{\mathcal{O}}, \delta \dot{r}^{\mathcal{O}} \right) \right\|_2$ and attitudinal~$\left\| \left( \delta q_{\D}^{\mathcal{O}} - q^I, \delta \omega_{\D}^{\mathcal{O}} \right) \right\|_2$ errors are shown in Figures~\ref{fig:transError} and~\ref{fig:rotError}, respectively.
Furthermore, the average thrust~$\| F^\D_d \|_2$ and torque~$\| \tau^\D \|_2$ inputs are shown in Figures~\ref{fig:thrust} and~\ref{fig:torque}.
We observe that as~$\jmax$ increases in Figures~\ref{fig:error}-~\ref{fig:attitude} that the deputy spacecraft achieves the docking configuration in a shorter time while incurring less error due to the fact more iterations of the optimization algorithm are allowed to occur.
A similar behavior can be seen in Figure~\ref{fig:avgCost} which shows the average cost across all 200 trials for all simulated values of~$\jmax$. This is due to the fact that the magnitude of the cost is directly related to the error, i.e.,~$\| z(k) - z_d \|_2$. Figure~\ref{fig:totAvgCost} shows the sum of the average cost from Figure~\ref{fig:avgCost} which demonstrates the difference in cost between different values of~$\jmax$.


\begin{figure}[H]
  \centering
  \begin{tabular}{cc}
    \begin{subfigure}{0.45\textwidth}
      \centering
      \includegraphics[width=\linewidth]{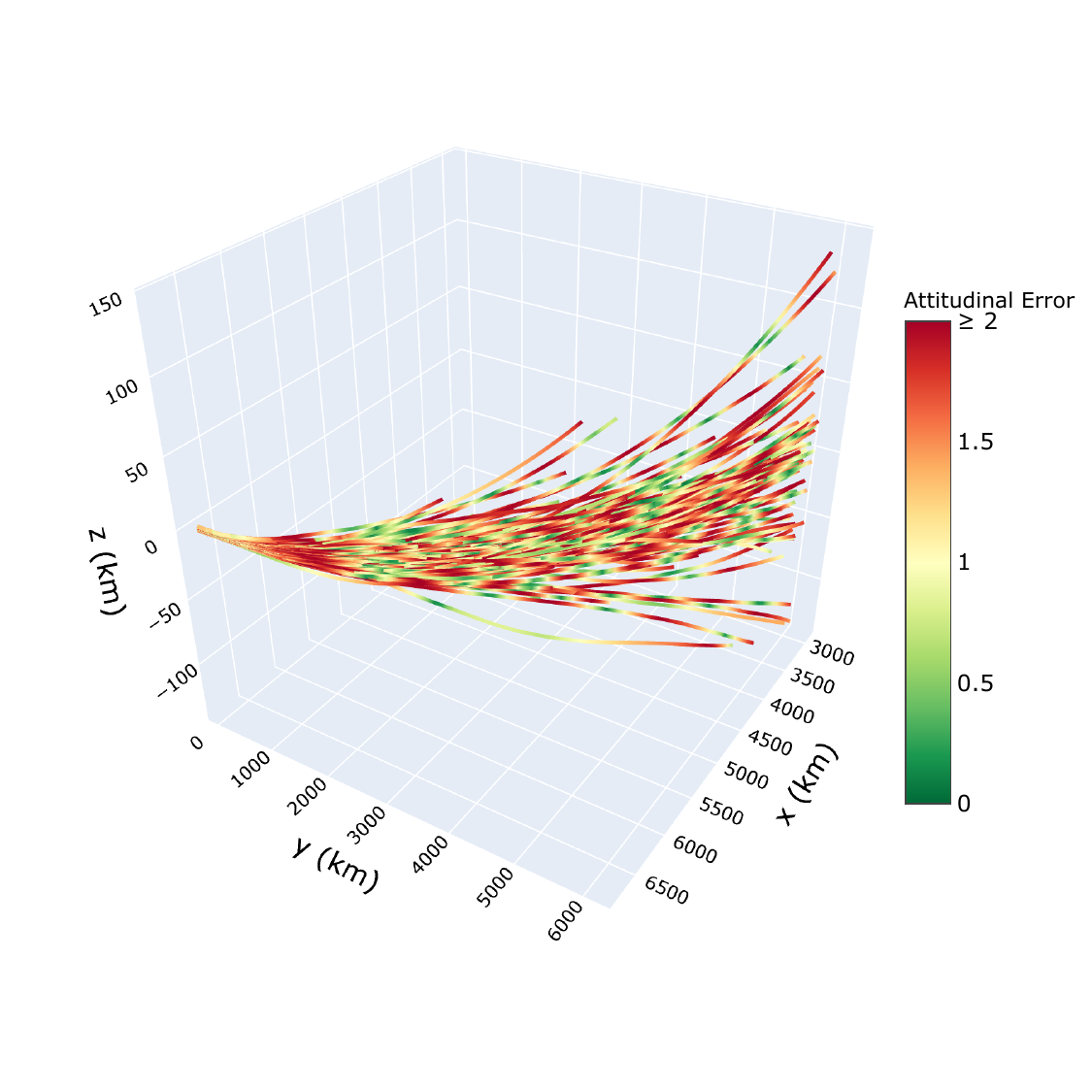}
      \caption{$j_{\max} = 1$}
      \label{fig:maxIter1}
    \end{subfigure} &
    \begin{subfigure}{0.45\textwidth}
      \centering
      \includegraphics[width=\linewidth]{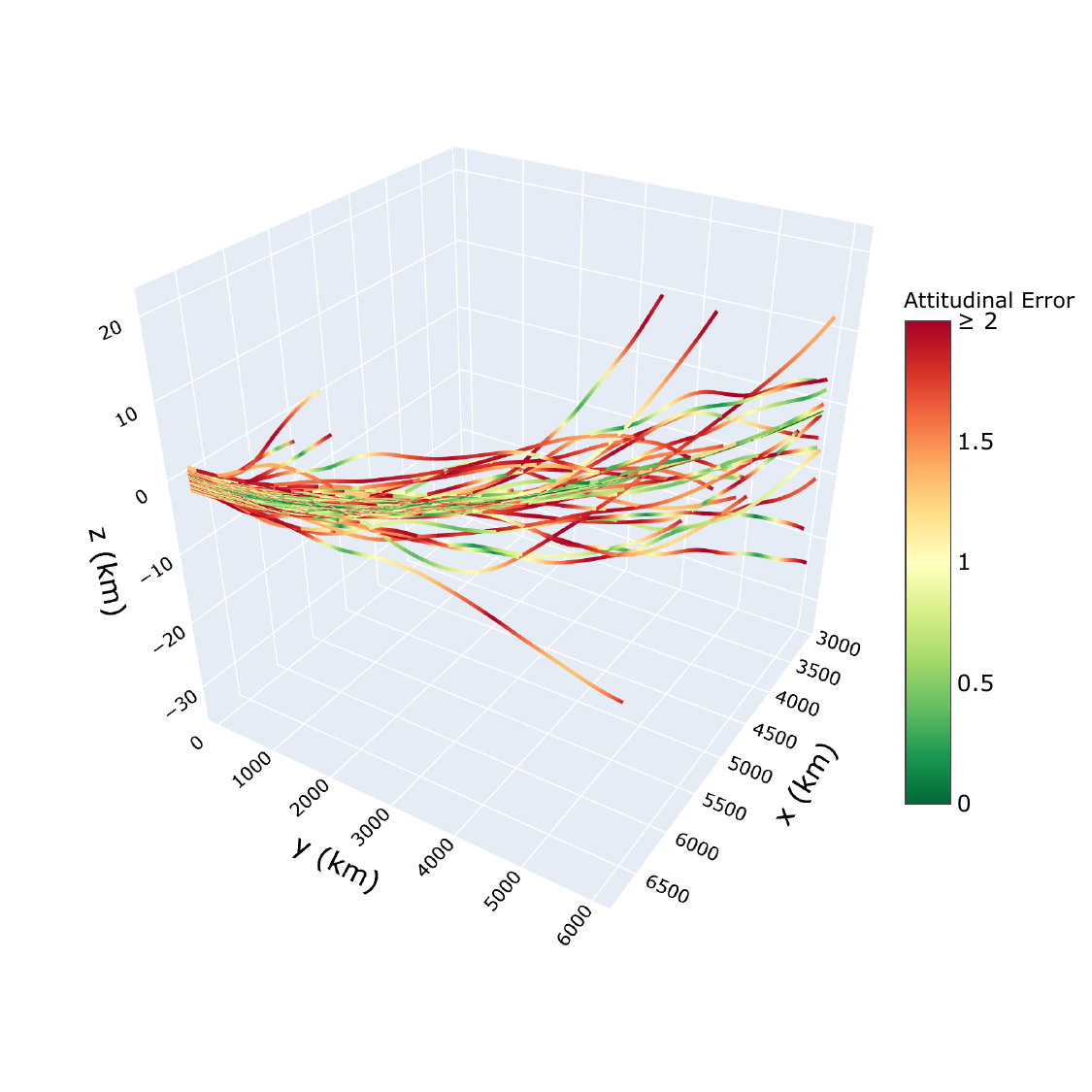}
      \caption{$j_{\max} = 2$}
      \label{fig:maxIter2}
    \end{subfigure}
  \end{tabular}
  
  \caption{Plot of the 3D deputy trajectories colored according to their attitudinal error for 200 initial conditions and~$\jmax=1,2$.}
  \label{fig:orbits1}
\end{figure}

\begin{figure}[H]
  \centering
  \begin{tabular}{cc}
    \begin{subfigure}{0.45\textwidth}
      \centering
      \includegraphics[width=\linewidth]{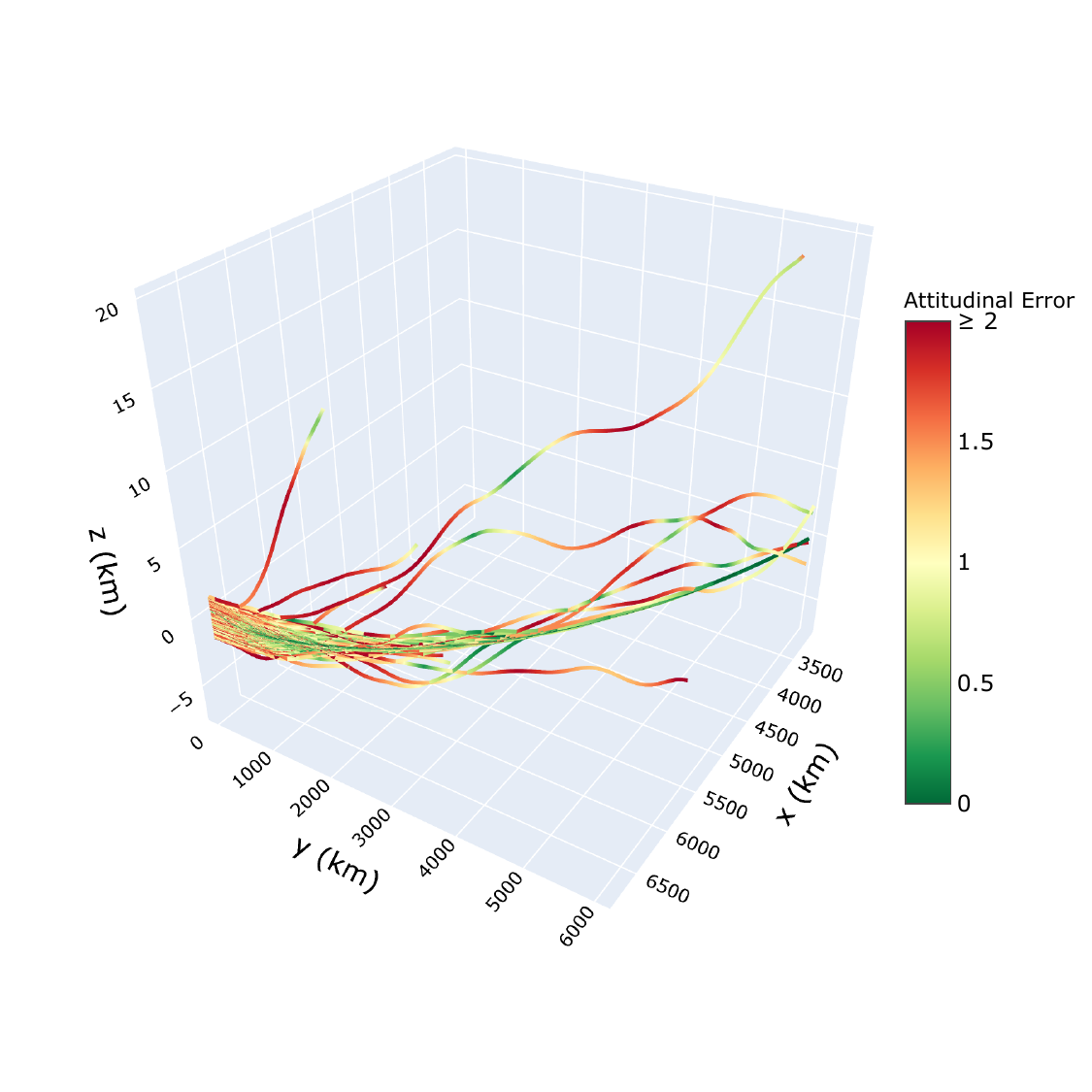}
      \caption{$j_{\max} = 3$}
      \label{fig:maxIter3}
    \end{subfigure} &
    \begin{subfigure}{0.45\textwidth}
      \centering
      \includegraphics[width=\linewidth]{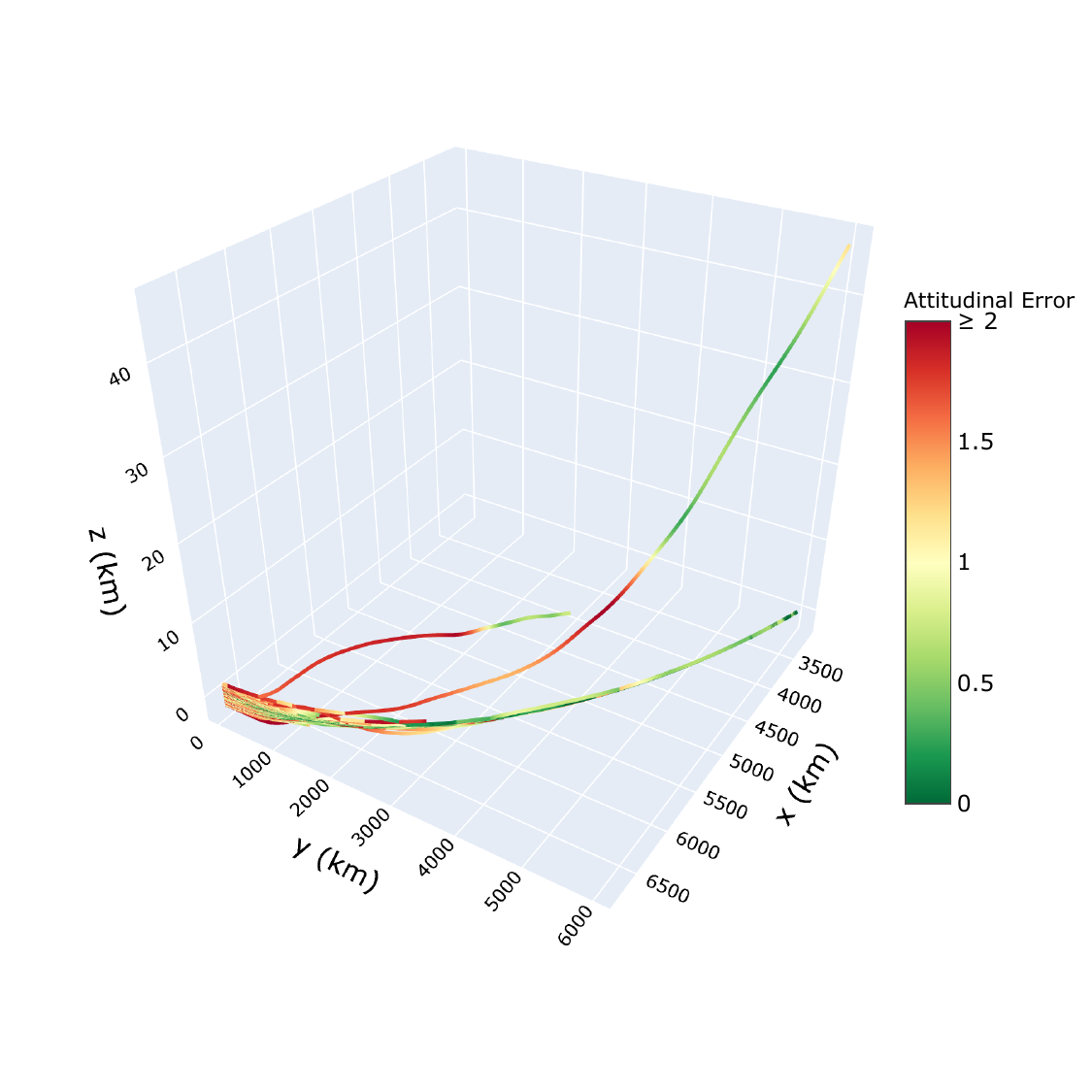}
      \caption{$j_{\max} = 4$}
      \label{fig:maxIter4}
    \end{subfigure} 
  \end{tabular}
  
  \caption{Plot of the 3D deputy trajectories colored according to their attitudinal error for 200 initial conditions and~$\jmax=3,4$.}
  \label{fig:orbits2}
\end{figure}

\begin{figure}[H]
  \centering
  \begin{tabular}{cc}
        \begin{subfigure}{0.45\textwidth}
      \centering
      \includegraphics[width=\linewidth]{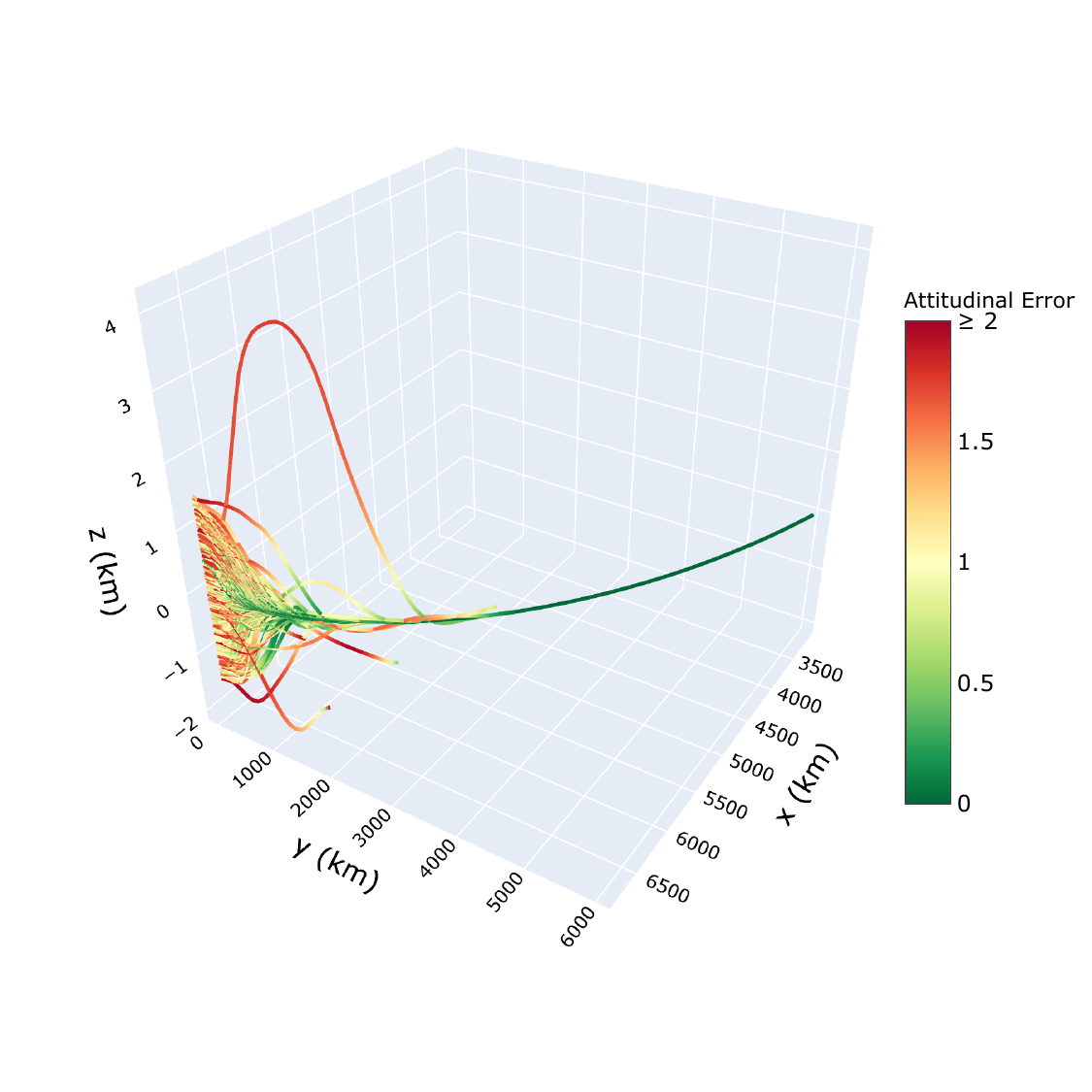}
      \caption{$j_{\max} = 5$}
      \label{fig:maxIter5}
    \end{subfigure} &
    \begin{subfigure}{0.45\textwidth}
      \centering
      \includegraphics[width=\linewidth]{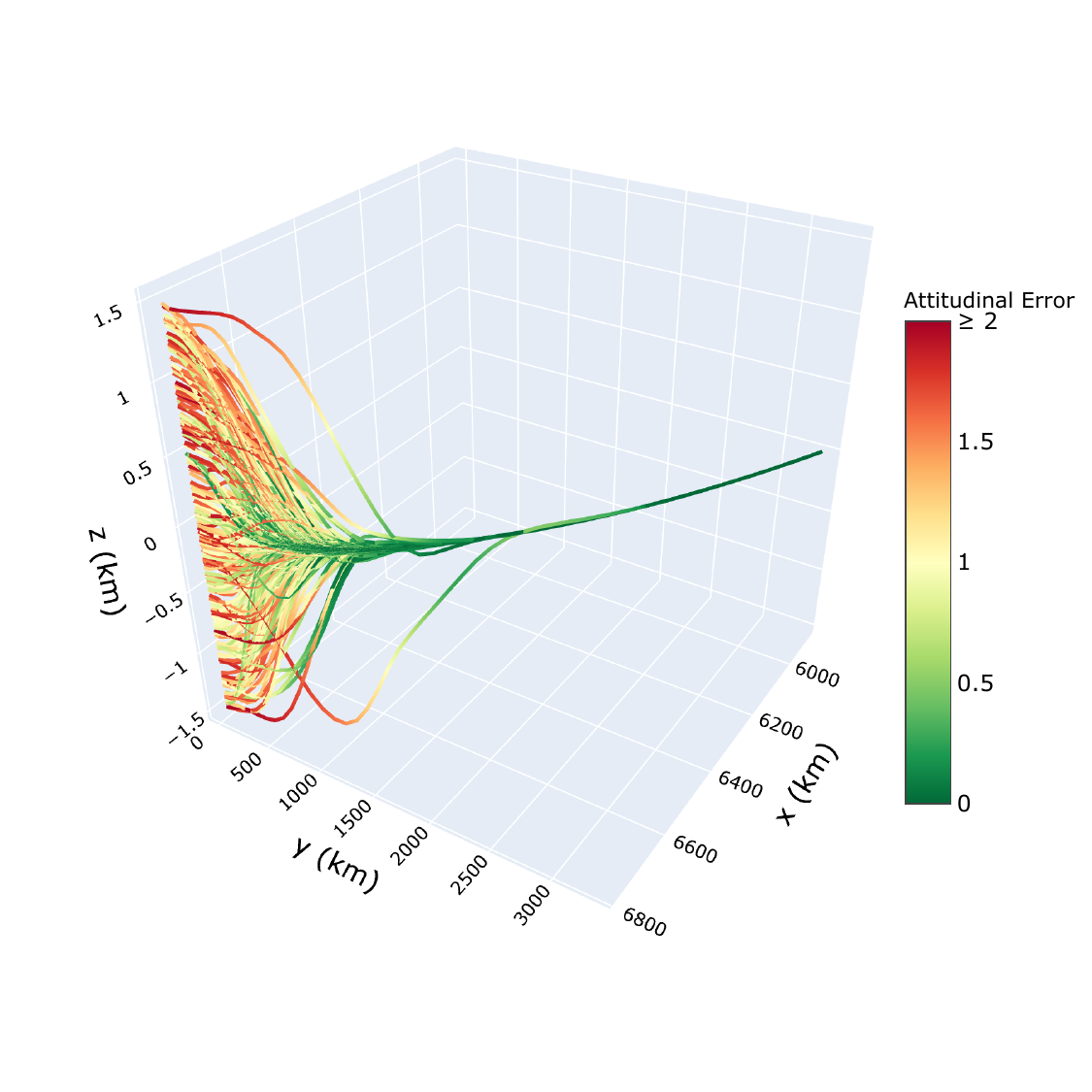}
      \caption{$j_{\max} = 6$}
      \label{fig:maxIter6}
    \end{subfigure} 
  \end{tabular}
  \caption{Plot of the 3D deputy trajectories colored according to their attitudinal error for 200 initial conditions and~$\jmax=5,6$.}
  \label{fig:orbits3}
\end{figure}

\begin{figure}[H]
  \centering
  \begin{tabular}{cc}
        \begin{subfigure}{0.45\textwidth}
      \centering
      \includegraphics[width=\linewidth]{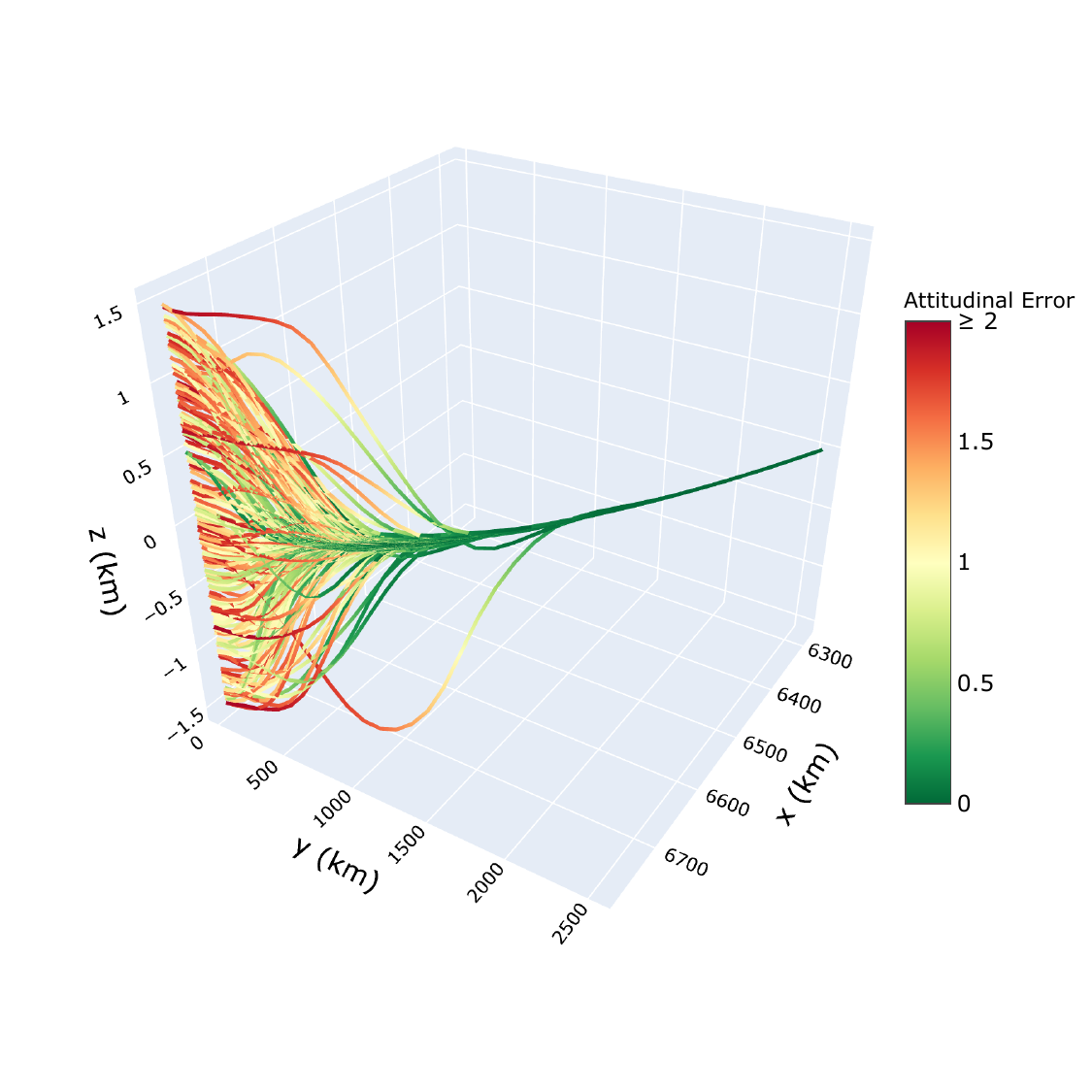}
      \caption{$j_{\max} = 7$}
      \label{fig:maxIter7}
    \end{subfigure} &
    \begin{subfigure}{0.45\textwidth}
      \centering
      \includegraphics[width=\linewidth]{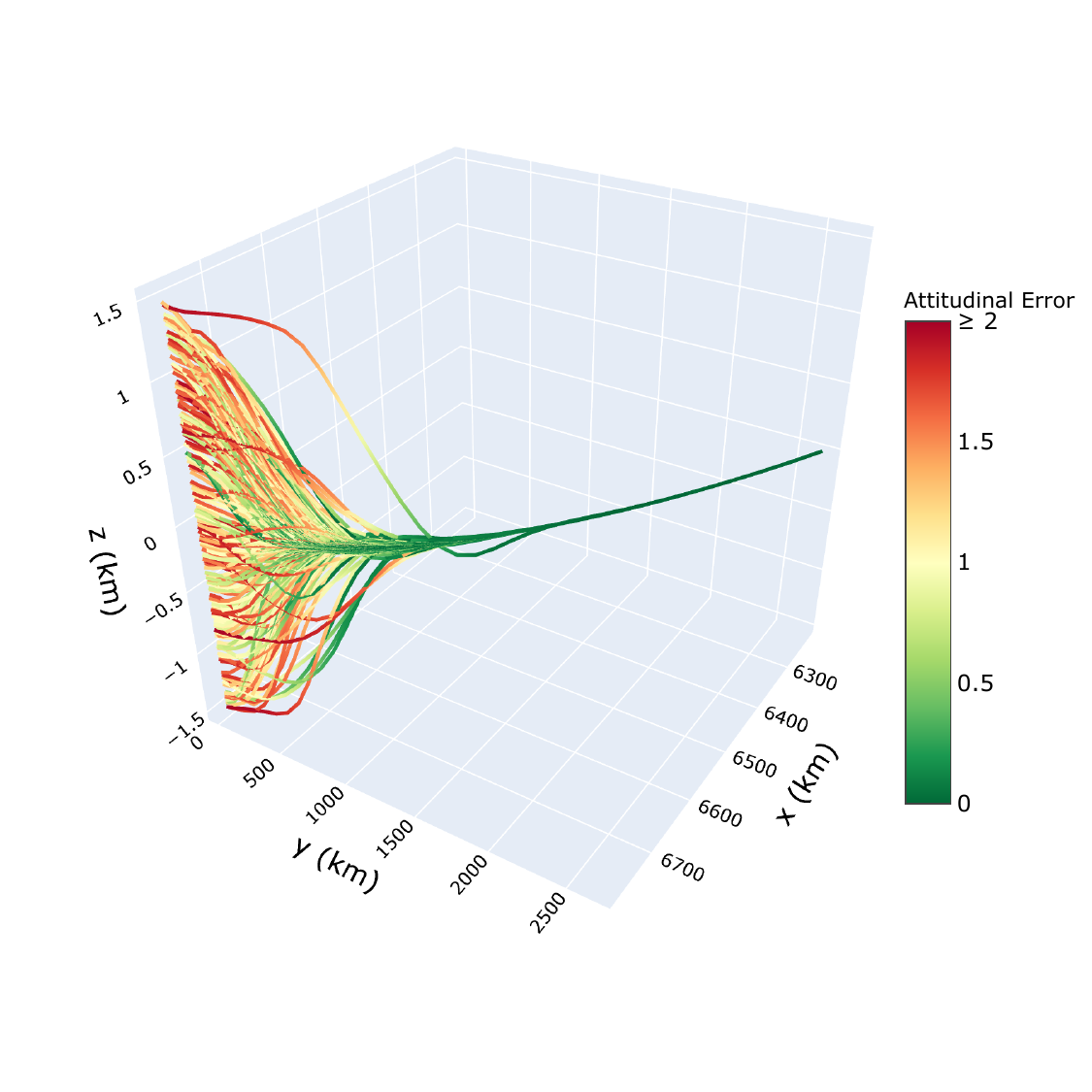}
      \caption{$j_{\max} = 8$}
      \label{fig:maxIter8}
    \end{subfigure} 
  \end{tabular}
  \caption{Plot of the 3D deputy trajectories colored according to their attitudinal error for 200 initial conditions and~$\jmax=7,8$.}
  \label{fig:orbits4}
\end{figure}

\begin{figure}[H]
  \centering
  \begin{tabular}{cc}
        \begin{subfigure}{0.45\textwidth}
      \centering
      \includegraphics[width=\linewidth]{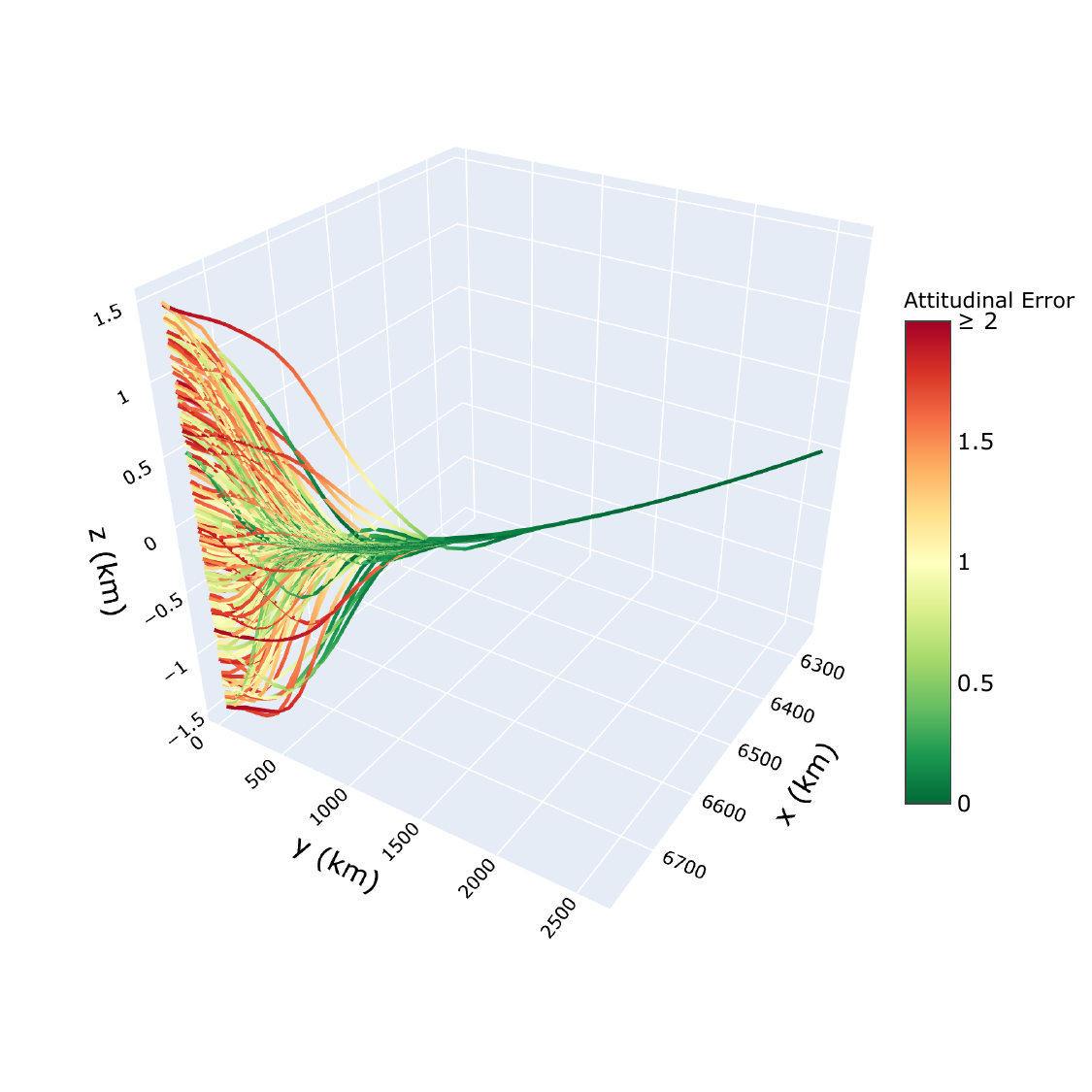}
      \caption{$j_{\max} = 9$}
      \label{fig:maxIter9}
    \end{subfigure} &
    \begin{subfigure}{0.45\textwidth}
      \centering
      \includegraphics[width=\linewidth]{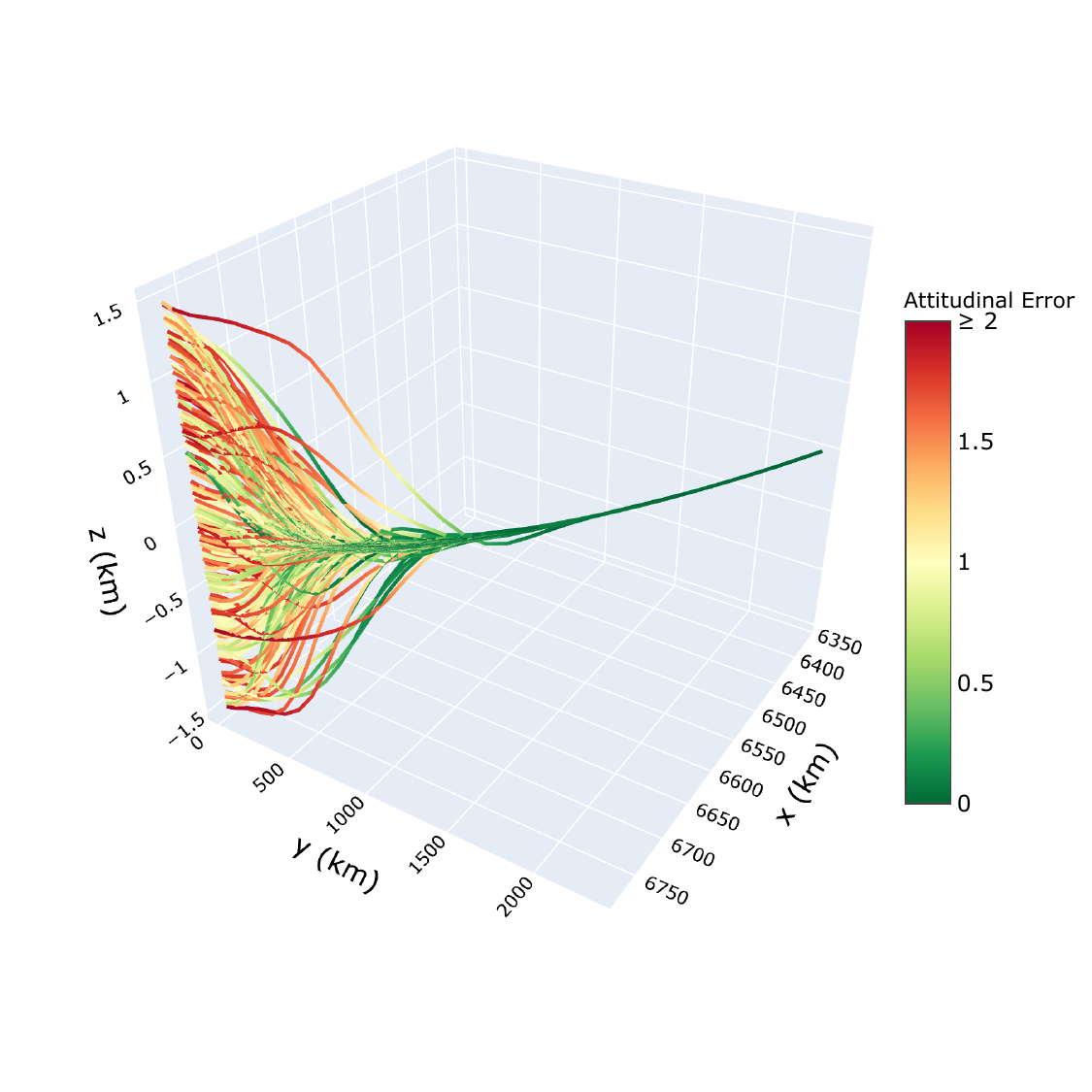}
      \caption{$j_{\max} = 10$}
      \label{fig:maxIter10}
    \end{subfigure} 
  \end{tabular}
  \caption{Plot of the 3D deputy trajectories colored according to their attitudinal error for 200 initial conditions and~$\jmax=9,10$.}
  \label{fig:orbits5}
\end{figure}

\begin{figure}[H]
  \centering
  \begin{tabular}{cc}
        \begin{subfigure}{0.45\textwidth}
      \centering
      \includegraphics[width=\linewidth]{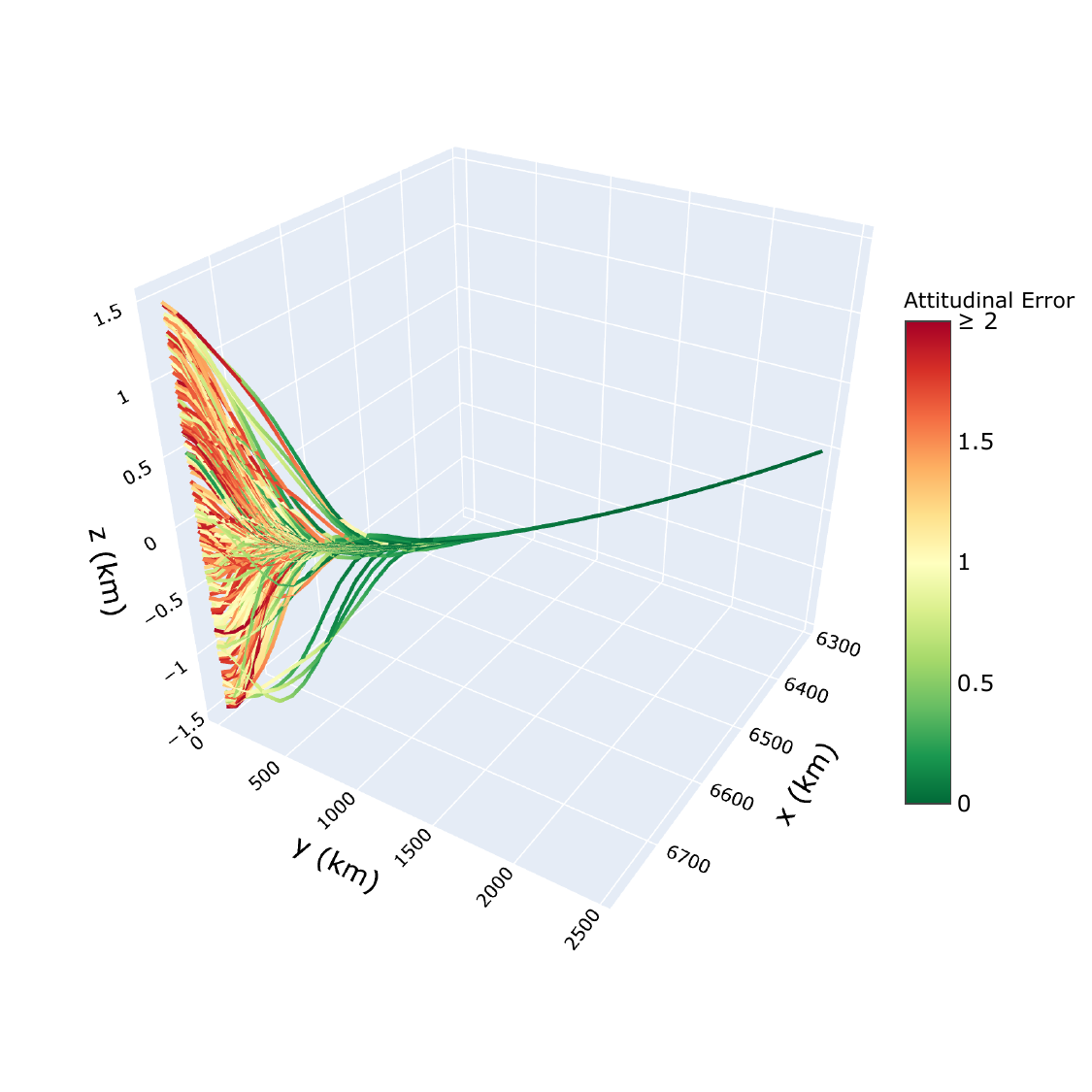}
      \caption{$j_{\max} = 50$}
      \label{fig:maxIter50}
    \end{subfigure} &
    \begin{subfigure}{0.45\textwidth}
      \centering
      \includegraphics[width=\linewidth]{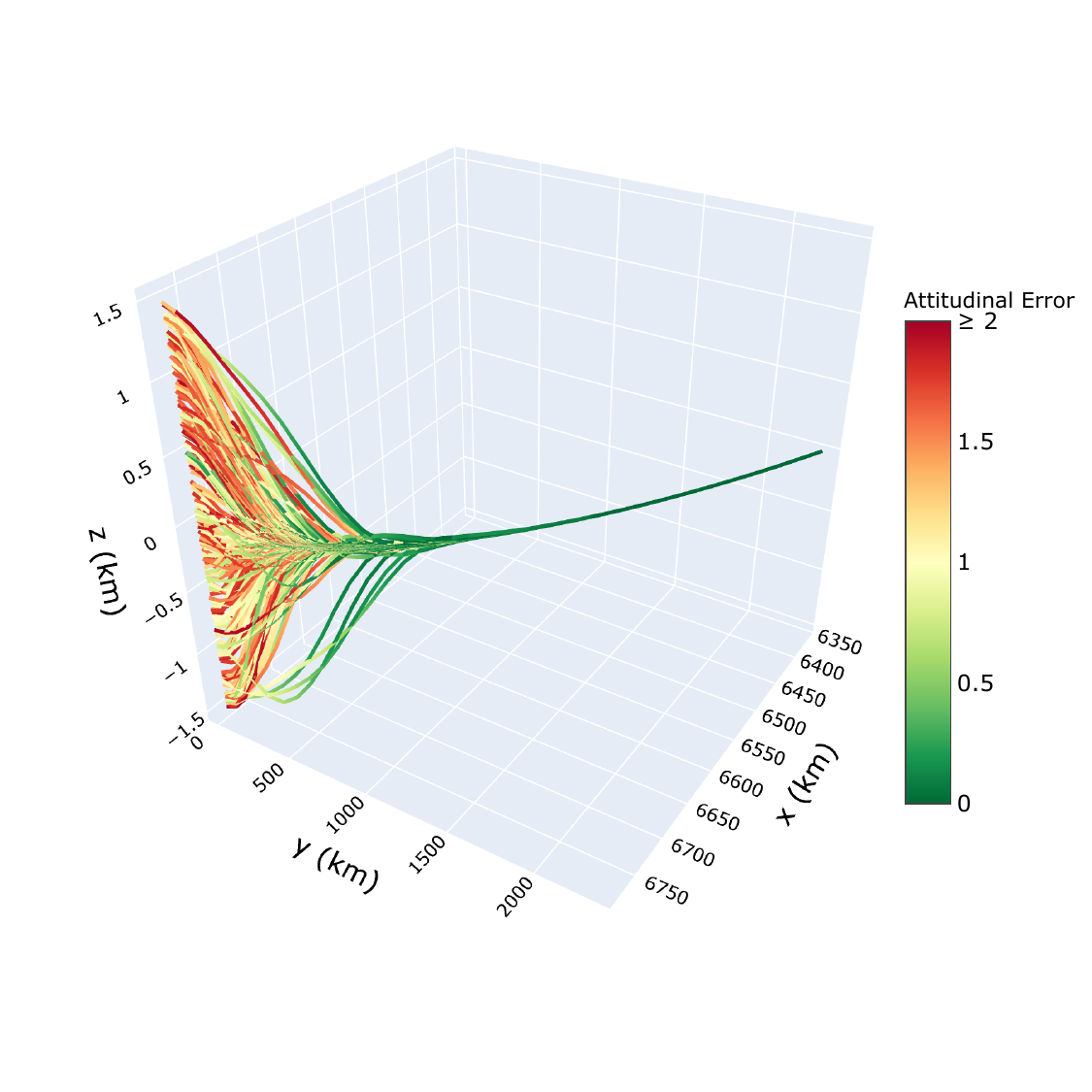}
      \caption{$j_{\max} = 100$}
      \label{fig:maxIter100}
    \end{subfigure} 
  \end{tabular}
  \caption{Plot of the 3D deputy trajectories colored according to their attitudinal error for 200 initial conditions and~$\jmax=50,100$.}
  \label{fig:orbits6}
\end{figure}

\begin{figure}[H]
  \centering
    \begin{subfigure}{0.45\textwidth}
      \centering
      \includegraphics[width=\linewidth]{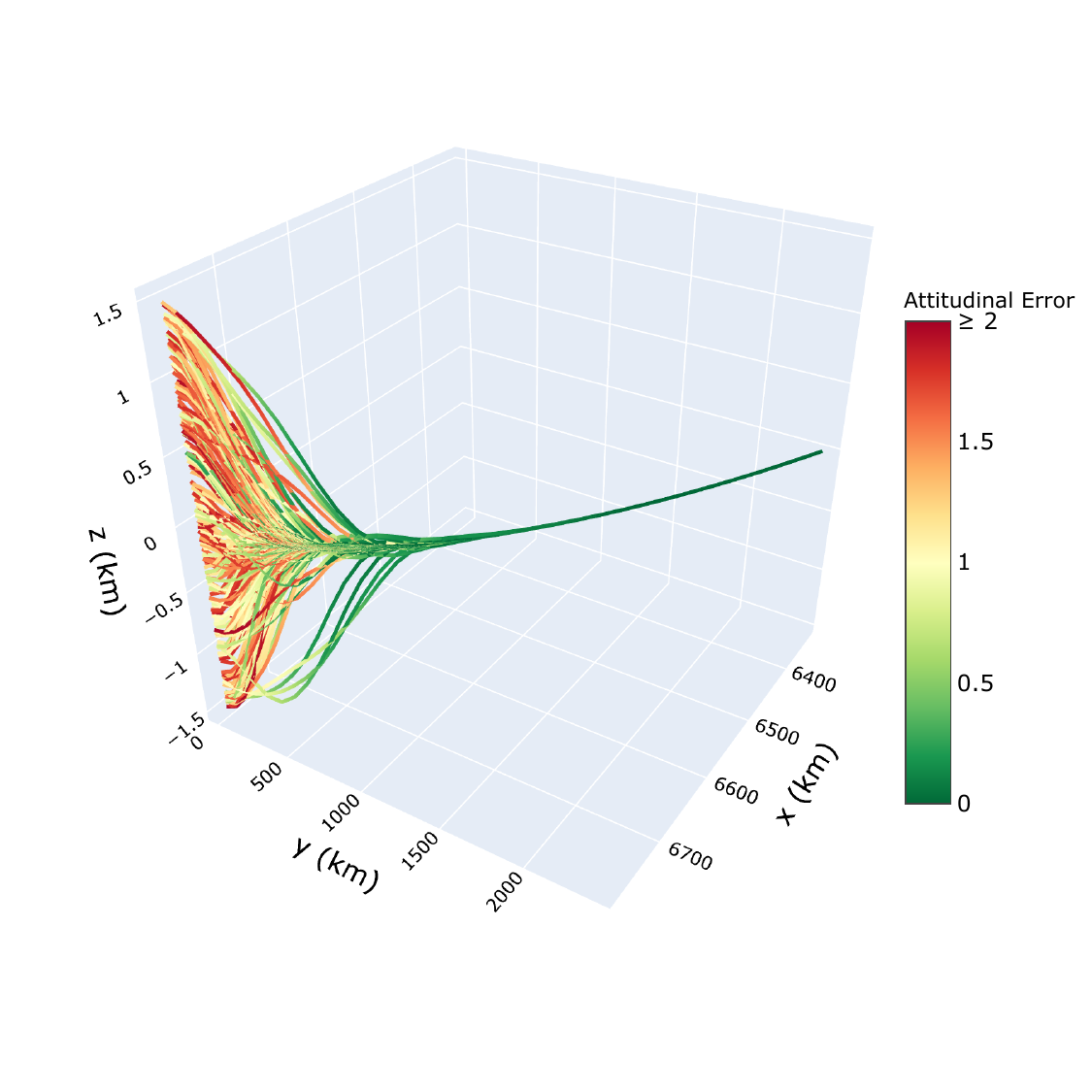}
      \caption{$j_{\max} = \text{Optimal}$}
    \end{subfigure} 
  \caption{Plot of the 3D deputy trajectories colored according to their attitudinal error for 200 initial conditions and~$\jmax=\text{Optimal}$.}
  \label{fig:orbits7}
\end{figure}

Figures~\ref{fig:orbits1}-~\ref{fig:orbits3} show the deputy's trajectories in 3D space using our time-constrained MPC algorithm for all 200 trials and all simulated values of~$\jmax$. These trajectories are colored based on the trial's current attitudinal error, i.e.,~$\left\| \left( \delta q_{\D}^{\mathcal{O}} - q^I, \delta \omega_{\D}^{\mathcal{O}} \right) \right\|_2$. 
As the deputy approaches the attitudinal docking configuration, i.e.,~$\left\| \left( \delta q_{\D}^{\mathcal{O}} - q^I, \delta \omega_{\D}^{\mathcal{O}} \right) \right\|_2 \to 0$, the deputy's trajectories turns from red/yellow to green.
However, in Figures~\ref{fig:orbits1}-\ref{fig:maxIter5} there are many trials that do not reach the docking configuration. This is due to the fact that, often times, the maximum allowable iteration constraint is met in the solving of Problem~\ref{prob:arpodMPC} at each~$k$ for these values of~$\jmax$ which causes the deputy to apply sub-optimal control inputs. 
In Figure~\ref{fig:maxIter1}, zero trials achieve the docking configuration and the trajectories do not converge to the chief orbit because the optimization algorithm is only allowed 1 iteration at each~$k$. 
In Figures~\ref{fig:maxIter2}-\ref{fig:maxIter5}, more trials achieve the docking configuration as~$\jmax$ increases, but still not all 200 trials converge.
For all~$\jmax \geq 6$, all 200 trials achieve the docking configuration which can be seen in Figures~\ref{fig:maxIter6}-\ref{fig:orbits7}.

\begin{figure}[H]
  \centering
  \begin{tabular}{cc}
    \begin{subfigure}{0.45\textwidth}
      \centering
      \includegraphics[width=\linewidth]{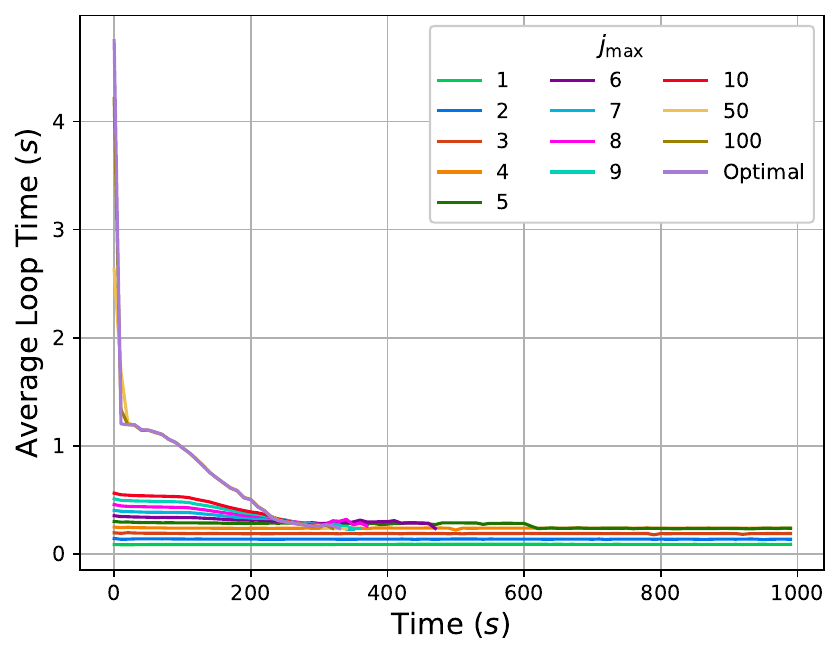}
      \caption{Average Loop Time}
      \label{fig:avgCPUTime}
    \end{subfigure} &
    \begin{subfigure}{0.45\textwidth}
      \centering
      \includegraphics[width=\linewidth]{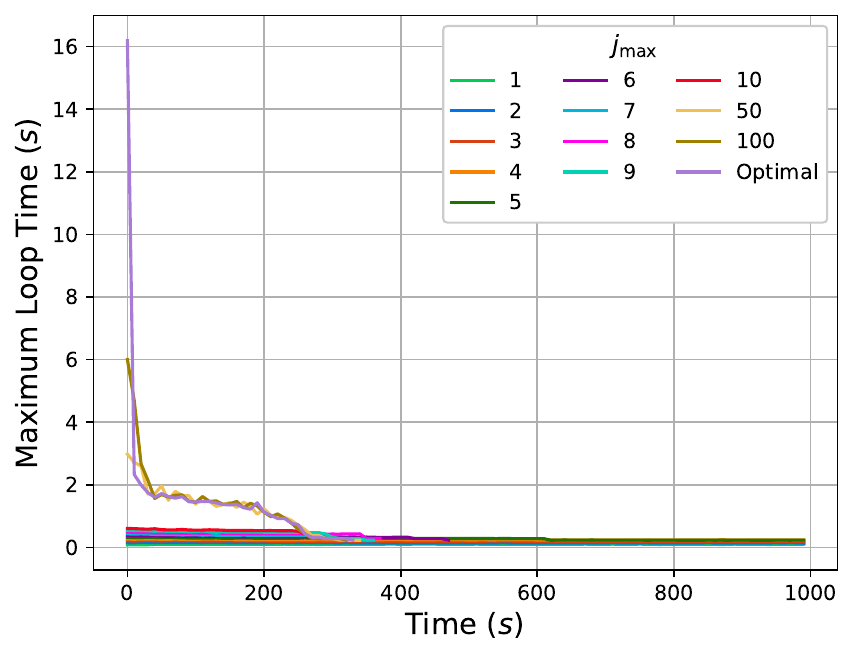}
      \caption{Maximum loop time}
      \label{fig:maxCPUTime}
    \end{subfigure} \\
    \begin{subfigure}{0.45\textwidth}
      \centering
      \includegraphics[width=\linewidth]{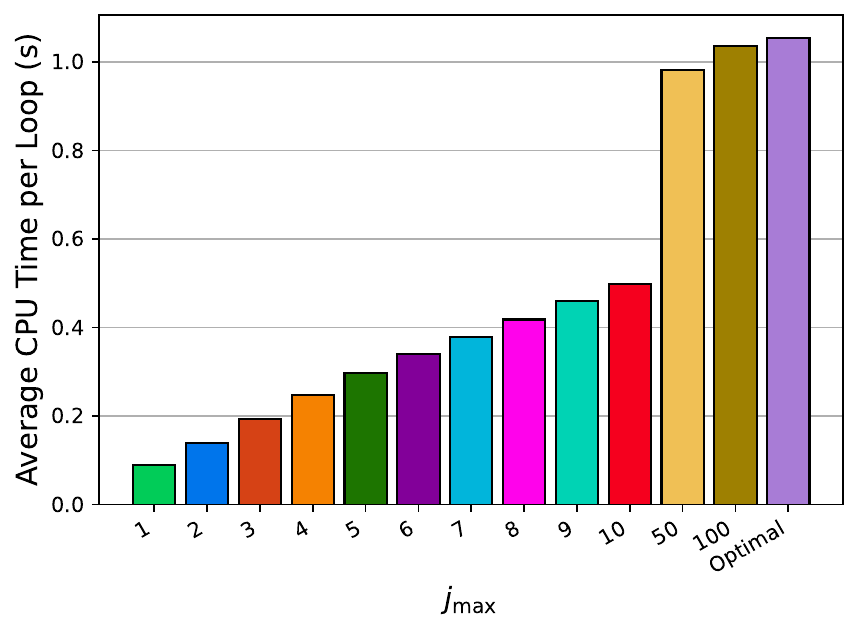}
      \caption{Average CPU Time Per Loop}
      \label{fig:avgCPUTimePerLoop}
    \end{subfigure}
    &
    \begin{subfigure}{0.45\textwidth}
      \centering
      \includegraphics[width=\linewidth]{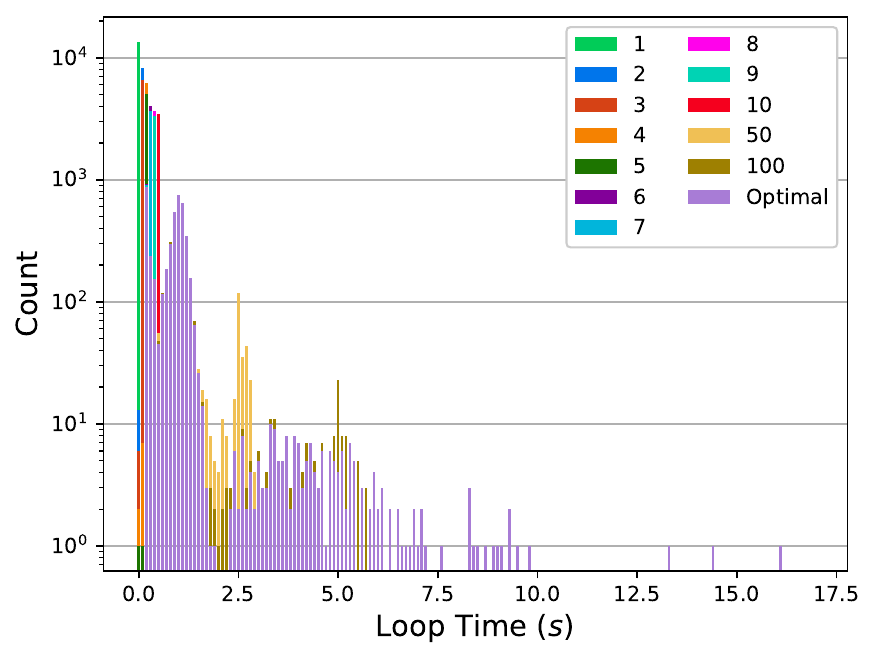}
      \caption{Timing Bins}
      \label{fig:timingBins}
    \end{subfigure} 
  \end{tabular}
  
  \caption{These plots display the average time per loop~\ref{fig:avgCPUTime}, maximum loop time~\ref{fig:maxCPUTime}, average CPU time per loop~\ref{fig:avgCPUTimePerLoop}, and the number of loops that fall into each timing bin~\ref{fig:timingBins}. As the maximum number of allowable iterations~$\jmax$ increases the time required to solve Problem~\ref{prob:arpodMPC} increases because the optimization algorithm is allowed to complete more iterations.}
  \label{fig:timing}
\end{figure}

Lastly, Figure~\ref{fig:timing} shows the timing statistics of the time-constrained MPC algorithm for all simulated values of~$\jmax$. Figures~\ref{fig:avgCPUTime} and~\ref{fig:maxCPUTime} show the average and maximum time the optimization algorithm spent solving for the control sequence~$\bm{u}(k)$ for all~$k$, respectively. 
The average loop time is the average time over all 200 trials the optimization algorithm spent solving Problem~\ref{prob:arpodMPC} for each~$k$ and the maximum loop time is the maximum time the optimization algorithm spent solving~\ref{prob:arpodMPC} at each~$k$.
The optimizer spends more time at the beginning of the problem solving for its control sequence~$\bm{u}(k)$ because the decision vector is initialized as~$\bm{u}_0 = \bm{0}_{6 \times N-1}$ and thereafter the previous iterates are used as the initial point for the optimization at the next time step also known as warm starting.
As~$\jmax$ increases the average and maximum loop time increases because the optimization algorithm is allowed to complete more iterations.
Figure~\ref{fig:avgCPUTimePerLoop} shows the average time the optimization algorithm takes to generate its control sequence for all values of~$\jmax$.

Figure~\ref{fig:timingBins} shows the number of loops that fall into each timing bin where the bin width is~$0.1$ seconds.
For~$\jmax \leq 10$ the optimization algorithm takes less than 2.5 seconds to solve Problem~\ref{prob:arpodMPC} for all 200 trials and for all~$k$.
Moreover, for~$\jmax=50,100$ the solving of Problem~\ref{prob:arpodMPC} takes less than 7.5 seconds. Lastly, for~$\jmax=\text{Optimal}$ there are 3 instances where the solving of Problem~\ref{prob:arpodMPC} takes more than 10 seconds which is longer than the sampling time~$t_s = 10$ seconds. 
Therefore, it may not be possible to implement conventional MPC onboard spacecraft to produce an optimal control sequence~$\bm{u}^*(k)$ for all~$k$.
However, we have shown that the deputy can achieve the docking configuration subject to computational time constraints by implementing time-constrained MPC for multiple values of~$\jmax$.

\section{Conclusion} \label{sec:conclusion}
We proposed a time-constrained MPC strategy for the 6 DOF ARPOD problem that accounts for nonlinear dynamics, robustness to perturbations, and computational time constraints.
We implemented time-constrained MPC on a SpaceCloud iX10-101 processing unit and showed that the deputy satellite can achieve the docking configuration while explicitly addressing computational time constraints.
We showed that utilizing time-constrained MPC allows us to greatly reduce the computation time for control inputs compared to conventional MPC.
Future research directions include considering additional constraints such as collision avoidance and line of sight constraints, and further exploring stability guarantees for time-constrained MPC.

\bibliography{references}

\end{document}